\documentclass[apj,twocolumn]{openjournal}

\usepackage{amsmath}
\usepackage{booktabs}
\usepackage{orcidlink}
\usepackage{color}

\newcommand{\fdir}{./}
\newcommand{\msun}{M_\odot}
\newcommand{\rsun}{R_\odot}

\newcommand{\fbg}{f_{\rm b,\ge 5\msun}}
\newcommand{\fbl}{f_{\rm b,< 5\msun}}
\newcommand{\sigmakccsn}{\sigma_{\rm k,ccsn}}
\begin{document}

\title{Compact Binary Formation in Open Star Clusters II: Difficulty
  of Gaia NS formation in low-mass star clusters}

\author{Ataru Tanikawa\orcidlink{0000-0002-8461-5517}$^{1}$}
\author{Long Wang\orcidlink{0000-0001-8713-0366}$^{2,3}$}
\author{Michiko S. Fujii\orcidlink{0000-0002-6465-2978}$^{4}$}

\affiliation{$^{1}$Center for Information Science, Fukui Prefectural
  University, 4-1-1 Matsuoka Kenjojima, Eiheiji-cho, Fukui 910-1195,
  Japan}
\affiliation{$^{2}$School of Physics and Astronomy, Sun Yat-sen
  University, Daxue Road, Zhuhai, 519082, China}
\affiliation{$^{3}$CSST Science Center for the Guangdong-Hong
  Kong-Macau Greater Bay Area, Zhuhai, 519082, China}
\affiliation{$^{4}$Department of Astronomy, Graduate School of
  Science, The University of Tokyo, 7-3-1 Hongo, Bunkyo-ku, Tokyo
  113-0033, Japan}
\email{Corresponding author: tanik@g.fpu.ac.jp}

\begin{abstract}

  Gaia mission offers opportunities to search for compact binaries not
  involved in binary interactions (hereafter inert compact binaries),
  and results in the discoveries of binaries containing one black hole
  (BH) or one neutron star (NS), called ``Gaia BHs'' and ``Gaia NSs'',
  respectively. We have assessed if Gaia BHs and NSs can be formed in
  open clusters through dynamical interactions. In order to obtain a
  large number of inert compact binaries similar to Gaia BHs and NSs,
  we have performed gravitational $N$-body simulations for a large
  number of open clusters whose total mass is $1.2 \times 10^8
  M_\odot$. These clusters have various masses, metallicities,
  densities, and binary fractions. We have found that open clusters
  form Gaia BHs ($10^{-6}$-$10^{-5} M_\odot^{-1}$) much more
  efficiently than Gaia NSs ($\lesssim 10^{-7} M_\odot^{-1}$) for any
  cluster parameters. This is quite inconsistent with observational
  results, because the reported numbers of Gaia BHs and NSs are
  $3$ and $21$, respectively. Additionally, we have
  switched off NS natal kicks for $10^4$ open clusters each weighing
  $10^3 M_\odot$ in order to retain a large number of NSs in open
  clusters. Then, open clusters form inert NS binaries originating
  from primordial binaries rather than formed through dynamical
  interactions. This means that Gaia NSs are formed dominantly on
  isolated fields, not in open clusters, if there is no NS natal
  kick. We have concluded that Gaia BHs can be dominantly formed in
  open clusters, however Gaia NSs cannot.

\end{abstract}

\maketitle

\section{Introduction}
\label{sec:Introduction}

Black holes (BHs) and neutron stars (NSs) are compact objects left
behind after the deaths of massive stars. They are clues to physical
processes involved in massive star evolution, and experimental sites
of strong gravitational fields. Thus, they have been explored
vigorously. Because of their darkness, only active BHs and NSs have
been discovered until quite recently. Active compact objects can be
X-ray binaries \citep[see][for review]{2017hsn..book.1499C},
gravitational wave sources \citep{2019PhRvX...9c1040A,
  2021PhRvX..11b1053A, 2023PhRvX..13a1048A}, and pulsars
\citep[see][for review]{2022ARA&A..60..495P}.

Very recently, inactive compact objects have started to be
discovered. Mainly, they are binary members, because they are found
through binary motions. They are called ``non-interacting'',
``detached'', ``dormant'', or ``inert'' compact binaries. Hereafter,
we unify the notation with inert compact binaries throughout this
paper. Such inert compact binaries have been discovered by
spectroscopic observations \citep{2018MNRAS.475L..15G,
  2022NatAs...6.1085S}, and by astrometric observations
\citep{2023MNRAS.518.1057E, 2023MNRAS.521.4323E, 2023ApJ...946...79T,
  2023AJ....166....6C} with the help of Gaia DR3
(\citealt{2023A&A...674A...1G, 2023A&A...674A..34G}; see review by
\citealt{2024arXiv240312146E}). Gaia DR3 seems to contain more inert
compact binaries according to several candidate lists
\citep{2022arXiv220700680A, 2023MNRAS.518.2991S, 2023arXiv230915143S,
  2023MNRAS.521.5927J, 2024arXiv240109531R}. Although other inert
compact binaries have been reported \citep{2019Natur.575..618L,
  2019Sci...366..637T, 2020A&A...637L...3R, 2021MNRAS.504.2577J,
  2022MNRAS.516.5945J, 2022A&A...665A.180L, 2022MNRAS.511.2914S},
their discoveries have been called into question
\citep{2020Natur.580E..11A, 2020A&A...641A..43B, 2020MNRAS.493L..22E,
  2021MNRAS.502.3436E, 2020Sci...368.3282V, 2022MNRAS.511L..24E,
  2022MNRAS.511.3089E, 2022MNRAS.512.5620E}.  In addition to inert
compact binaries, several single BHs/NSs have been detected by
microlensing observations \citep{2022ApJ...933L..23L,
  2022ApJ...933...83S, howil2024uncovering}.

Gaia BH1 and BH2 (hereafter Gaia BHs), which are inert BH binaries
discovered from Gaia DR3, and named by \cite{2023MNRAS.518.1057E,
  2023MNRAS.521.4323E}, have two impacts. First, Gaia BHs have the
longest orbital periods among BH binaries discovered so far. In
particular, Gaia BH2 breaks the record of BH binary periods by 10
times. These discoveries give the impression that Gaia's astrometric
observations open up a new parameter space of BH binaries. Second,
Gaia BHs are hard to form through isolated binary evolution. If Gaia
BHs are formed through isolated binary evolution, common envelope
ejection should be 10 times more efficient than expected
(\citealt{2022ApJ...931..107C, 2023MNRAS.518.1057E,
  2023ApJ...953...52S}; but see
\citealt{2024arXiv240313579K}). However, a recent study has not
supported such efficient common envelope ejection for stars massive
enough to form BHs \citep{2022ApJ...937L..42H}. Triple stars
\citep{2023MNRAS.518.1057E, 2023MNRAS.521.4323E, 2023arXiv231203066G}
and open clusters \citep{2020PASJ...72...45S, 2023MNRAS.526..740R,
  2024MNRAS.527.4031T, 2024ApJ...965...22D} have been suggested as
formation sites of Gaia BHs. Because Gaia BHs are the Galactic disk
components, it is difficult to suppose that they are formed in
globular clusters and galactic centers.

During the review process of this paper, Gaia BH3 has been reported by
\cite{2024arXiv240410486G}. It could belong to the ED-2 stream, which
is possibly a tidally disrupted globular cluster
\citep{2024arXiv240411604B}. Thus, it would be formed through
dynamical interactions in the globular cluster
\citep{2024arXiv240413036M}, which is similar to Gaia BH formation in
open clusters \citep{2020PASJ...72...45S, 2023MNRAS.526..740R,
  2024MNRAS.527.4031T, 2024ApJ...965...22D}. However, Gaia BH3 could
be formed from isolated binaries \citep{2024arXiv240413047E,
  2024arXiv240417568I}, since it is a very metal-poor binary, [Fe/H]
$= -2.56$ \citep{2024arXiv240410486G}. The origin of Gaia BHs is an
open question.

\cite{2024OJAp....7E..27E} have discovered Gaia NS1 which is most
likely to be the first inert NS binary from Gaia DR3
\citep{2023A&A...674A..34G}. The NS candidate mass ($\sim 1.9 \msun$)
is higher than other inert NS binary candidates which could be a
massive white dwarfs (WDs) as described in the literature
\citep{2023A&A...677A..11G, 2023SCPMA..6629512Z}. Gaia NS1 is hard to
form through isolated binary evolution without efficient common
envelope ejection, similarly to Gaia BHs. \cite{2024OJAp....7E..27E}
have also mentioned that they have found about 20 more inert NS
binaries, called ``Gaia NSs'' hereafter. \cite{2024arXiv240500089E}
have published their detail information during the review process of
this paper. This should imply that Gaia NSs are formed more
efficiently than Gaia BHs for the following two reasons. First, the
number of Gaia NSs discovered is larger than that of Gaia BHs
discovered. Second, Gaia NSs is harder to detect than Gaia BHs,
because NSs indicate smaller astrometric signals than BHs; NSs swing
around their companions less strongly than BHs due to their smaller
masses. Another importance of Gaia NS1 is its eccentricity. The
eccentricity is $\sim 0.1$ \citep{2024OJAp....7E..27E}, while Gaia BH1
and BH2 have $\sim 0.5$ \citep{2023MNRAS.518.1057E,
  2023MNRAS.521.4323E}. The eccentricity of Gaia NS1 may be too small
to be formed through dynamical interactions. We also look into the
eccentricities of all the Gaia NSs.

In this paper, we test if open clusters can form Gaia NSs more
efficiently than Gaia BHs by means of gravitational $N$-body
simulations. We have found that the formation rate of Gaia NSs is much
smaller than that of Gaia BHs if we adopt the conventional single and
binary star models. Open clusters cannot retain enough NSs to form
Gaia NSs because of their large natal kicks. In order to retain a
sufficiently large number of Gaia NSs in open clusters, we switch off
natal kicks of NSs, although this assumption is
unrealistic. Nevertheless, the formation rate of Gaia NSs is still
comparable to that of Gaia BHs, in contrast to the observational
implication.

The structure of this paper is as follows. We describe our simulation
method in section \ref{sec:Method}. We show our results in section
\ref{sec:Results}. We discuss if Gaia NSs can be formed in open
clusters in section \ref{sec:Discussion}. We summarize this paper in
section \ref{sec:Summary}.

\section{Method}
\label{sec:Method}

We use an $N$-body code PETAR \citep{2020MNRAS.497..536W}, which is
based on the particle-tree and particle-particle algorithm
\citep{2011PASJ...63..881O, 2017PASJ...69...81I}, and highly
parallelized by Framework for Developing Particle Simulator with
Massage Passing Interface, OpenMP, and Single Instruction Multiple
Data \citep[FDPS;][]{2016PASJ...68...54I, 2020PASJ...72...13I}. The
slow-down algorithmic regularization method
\citep[SDAR:][]{2020MNRAS.493.3398W} treats binary orbits and close
encounters in order to keep high accuracy efficiently. The Galactic
potential is modeled by GALPY \citep{2015ApJS..216...29B}. Single and
binary star evolutions are implemented with the BSE code
\citep{2000MNRAS.315..543H, 2002MNRAS.329..897H,
  2020A&A...639A..41B}. Common envelope evolution is the most
important process to form compact binaries. The BSE code deals with
common envelope evolution, adopting the $\alpha$ formalism
\citep{1984ApJ...277..355W}. We adopt $\alpha_{\rm CE}=1$ and
$\lambda_{\rm CE}$ of \cite{2014A&A...563A..83C}. The choice of
$\alpha_{\rm CE}$ prevents inert compact binaries from being formed
from primordial binaries.

NS population retained in open clusters largely depends on NS natal
kicks. Here, we remark on NS natal kicks in our model. NSs receive
kick velocities at their births because of asymmetric supernova
explosions. The kick velocities are quite different between NSs born
from electron capture (EC) and core collapse (CC) supernovae (SNe). In
our model, a star causes a CCSN when its carbon-oxygen (CO) core
exceeds both of the Chandrasekhar mass limit ($1.44\msun$ here) and
$0.773 M_{\rm c,he}-0.35 \msun$, where $M_{\rm c,he}$ is the He core
mass of the star. NSs left behind CCSNe have kick velocity
distribution given by a single Maxwellian with $\sigmakccsn=265$
km~s$^{-1}$ \citep{2005MNRAS.360..974H}. If the NSs acquires on
fallback mass, the kick velocities are reduced by $1-f_{\rm
  fallback}$, where $f_{\rm fallback}$ is the fraction of the fallback
mass \citep{2012ApJ...749...91F}. This prescription is also applied to
BH natal kicks. ECSNe \citep{1980PASJ...32..303M, 1984ApJ...277..791N,
  1987ApJ...322..206N, 2007AIPC..924..598V} are thought to raise much
smaller kick velocities than CCSNe because of slightly asymmetric
explosion \citep{2013ApJ...772..150J, 2018ApJ...865...61G}. In our
model, a star generates an ECSN when it does not cause a CCSN, and its
CO core mass is $\ge 1.6 \msun$. Their natal kick velocities are
expressed by a single Maxwellian with $3$ km~s$^{-1}$
\citep{2004ApJ...612.1044P, 2018ApJ...865...61G}. About $1$ \% of NSs
are formed from ECSNe under the initial mass function we adopt.

NSs with small natal kicks can be formed from accreting WDs
\citep{1987ApJ...320..304B, 1991ApJ...367L..19N, 1992ApJ...391..228W,
  1999ApJ...516..892F, 2006ApJ...644.1063D, 2010PhRvD..81d4012A,
  2023arXiv230617381M, 2023MNRAS.525.6359L}. This channel is called
accretion-induced collapse (AIC). Such NSs could be inside of globular
clusters \citep{2008MNRAS.386..553I, 2023MNRAS.525L..22K}. In our
simulations, this channel is achieved from two paths. First, an
oxygen-neon (ONe) WD increases its mass to the Chandrasekhar mass
limit, accreting materials through mass transfer from its MS or
post-MS (PMS) companion (non-degenerate companion). Second, two WDs
merge without explosion. We describe the second path, since it is not
implemented in previous BSE codes. We suppose that WD mergers create
NSs under the following conditions.
\begin{enumerate}
\item The total mass of two WDs exceeds the Chandrasekhar mass limit.
\item Neither of two WDs is a HeWD. Such WD merger products are
  thought to explode due to the double detonation scenario
  \citep{1982ApJ...257..780N, 1986ApJ...301..601W} in double WD
  systems \citep{2010Natur.465..322P, 2010ApJ...709L..64G,
    2013ApJ...770L...8P, 2021MNRAS.503.4734P, 2018ApJ...865...15S,
    2018ApJ...868...90T, 2019ApJ...885..103T, 2020A&A...635A.169G,
    2022ApJ...930...92F, 2023ApJ...944...22Z, 2023OJAp....6E..28E}.
\item Both of two WDs do not consist of two carbon-oxygen (CO) WDs
  with $\ge 0.8 \msun$.  This type of merger products can experience
  explosion due to the violent merger scenario
  \citep{2010Natur.463...61P, 2011A&A...528A.117P,
    2012MNRAS.424.2222P, 2012ApJ...747L..10P}. This condition is
  referred from \cite{2015ApJ...807..105S, 2016ApJ...821...67S}. Note
  that this criterion naturally includes an explosion due to the
  spiral instability \citep{2015ApJ...800L...7K}.
\end{enumerate}
These conditions are motivated by binary population synthesis modeling
of type Ia SNe from WD mergers \citep{2012A&A...546A..70T,
  Ruiter_2019}. We assume that two WD merger creates a NS with
the following properties.
\begin{enumerate}
\item No natal kick. AIC develops only small asymmetry, similarly to
  an ECSN \citep{2006ApJ...644.1063D, 2015MNRAS.453.1910S}.
\item The same mass as the total mass of the two WDs. WD mergers
  trigger nuclear reactions, however eject at most $0.1 \msun$
  \citep{2014MNRAS.438...14D, 2018ApJ...869..140K}. Although the
  system can eject its mass during AIC, the ejecta mass is at most
  $\sim 0.01 \msun$ \citep{2006ApJ...644.1063D, 2023arXiv230617381M,
    2023MNRAS.525.6359L}. The total mass of a WD merger product
  ($\gtrsim 1.4 \msun$) is larger than mass loss from the WD merger to
  AIC ($\sim 0.1 \msun$). Thus, we assume that the system loses no
  mass from the merger to AIC.
\end{enumerate}

In Table \ref{tab:SummaryOfModels}, we summarize our cluster
models. In the fiducial model, the initial cluster mass ($M$) and
stellar metallicity ($Z$) are $1000\msun$ and $0.02$, respectively. A
cluster is on a circular orbit at the distance of $8$ kpc from the
Galactic center.  The initial cluster mass density $\rho$ is
$20\msun~{\rm pc}^{-3}$, where $\rho$ is defined as mass density
inside the half-mass radius of a cluster. The initial binary fraction
depends on primary star masses ($m_1$). The initial binary fraction
for $m_1 \ge 5\msun$ ($\fbg$) is unity (or $100$ \%), while the
initial binary fraction for $m_1 < 5 \msun$ ($\fbl$) is $0.2$. Note
that the initial binary fraction is defined as $N_{\rm b}/(N_{\rm
  s}+N_{\rm b})$, where $N_{\rm s}$ and $N_{\rm b}$ are the numbers of
single and binary stars, respectively. Single stars and primary stars
in binaries have Kroupa's initial mass function
\citep{2001MNRAS.322..231K} in the range from $0.08 \msun$ to $150
\msun$. When the primary star of a binary has more than or equal to
$5\msun$, the binary mass ratio ($q$), orbital period ($P$), and
orbital eccentricity (e) distributions are subject to initial
conditions of \cite{2012Sci...337..444S}: $f(q) \propto q^{-0.1} \;
(0.1 \le q \le 1)$, $f(\log P) \propto (\log P)^{-0.55} \; (0.15 <
\log (P/{\rm day}) < 5.5$, and $f(e) \propto e^{-0.42} \; (0 \le e \le
1)$. When the primary star of a binary has less than $5\msun$, the
binary mass ratio ($q$), orbital period ($P$), and orbital
eccentricity (e) distributions are subject to initial conditions of
\cite{1995MNRAS.277.1491K, 1995MNRAS.277.1507K} with modification of
\cite{2017MNRAS.471.2812B}: $f(q) \propto {\rm const}$, $f(\log P)
\propto (\log(P/{\rm day})-1)/(45+[\log(P/{\rm day})-1]^2)$ $(10 \le
P/{\rm day} \le 10^{8.43})$, and $f(e) \propto e$.  We make These
initial conditions by means of MCLUSTER
\citep{2011MNRAS.417.2300K}. The simulations are finished at $1$ Gyr.

\begin{table}
  \centering
  \caption{Summary of models. Parameters indicated by ``--'' are the
    same as those of the fiducial model.} \label{tab:SummaryOfModels}
  \begin{tabular}{lccccc}
    \hline
    Name & $M$       & $Z$ & $\rho$ & $\fbl$ & $\sigmakccsn$ \\
         & [$\msun$] &     & [$\msun{\rm pc}^{-3}$] & & [kms$^{-1}$] \\
    \hline
    Fiducial      & $1000$ & $0.02$   & $20$  & $0.2$ & $265$ \\
    $M=200\msun$  & $200$  & --       & --    & --    & -- \\
    $M=500\msun$  & $500$  & --       & --    & --    & -- \\
    $M=2000\msun$ & $2000$ & --       & --    & --    & -- \\
    $Z=0.01 $     & --     & $0.01$   & --    & --    & -- \\
    $Z=0.005$     & --     & $0.005$  & --    & --    & -- \\
    $Z=0.002$     & --     & $0.002$  & --    & --    & -- \\
    $Z=0.0002$    & --     & $0.0002$ & --    & --    & -- \\
    $\rho=2$      & --     & --       & $2$   & --    & -- \\
    $\rho=200$    & --     & --       & $200$ & --    & -- \\
    $\fbl=0$       & --     & --       & --    & $0$   & -- \\
    $\fbl=0.5$     & --     & --       & --    & $0.5$ & -- \\
    No-kick       & --     & $0.002$  & --    & --    & $0$ \\
    \hline
  \end{tabular}
\end{table}

We prepare cluster models in which one parameter is different from
those in the fiducial model. The $M=200\msun$, $M=500\msun$ and
$M=2000\msun$ models have clusters with the initial masses of
$200\msun$, $500\msun$ and $2000\msun$, respectively. The $Z=0.01$,
$Z=0.005$, $Z=0.002$, and $Z=0.0002$ models have clusters with stellar
metallicities of $Z=0.01$, $0.005$, $0.002$, and $0.0002$
respectively. The $\rho=2$ and $\rho=200$ models have clusters with
the initial mass densities of $2$ and $200\msun~{\rm pc}^{-3}$,
respectively. The $\fbl=0$ and $\fbl=0.5$ models have $\fbl=0$ and
$0.5$, respectively. Note that $\fbg$ is fixed to unity, which is
reasonable according to \cite{2012Sci...337..444S} and
\cite{2017ApJS..230...15M}.

Additionally, we make the no-kick model in which no CCSNe generate
kick velocities, in order to investigate effects of CCSN natal
kicks. We set the metallicity to $Z=0.002$, because the visible star
of Gaia NS1 has small metallicity, $[{\rm Fe/H}] = -1.23$
\citep{2024OJAp....7E..27E}.

We generate a large number of open clusters with different
realizations, so that each model contains $10^7 \msun$ open clusters
in total. The total mass of simulated clusters is $1.3 \times 10^8
\msun$. We can expect that we obtain $\sim 100$ Gaia BHs from each
cluster model, because the formation efficiency of Gaia BHs is $\sim
10^{-6}M_\odot^{-1}$ \citep{2023MNRAS.526..740R, 2024MNRAS.527.4031T,
  2024ApJ...965...22D}. This enables us to make comparison between the
formation efficiencies of Gaia BHs and NSs.

From our simulations, we pick up compact binaries with the following
criteria:
\begin{enumerate}
\item $10^2 \le P/{\rm day} \le 10^4$, where $P$ is a binary orbital
  period. This criterion may look loose, since the orbital periods of
  Gaia BHs and Gaia NSs are $100$-$4000$ days, and $100-1000$ days,
  respectively. Thus, we also investigate the case for $10^2 \le
  P/{\rm day} \le 10^3$.
\item $m_2 \le 1.1 \msun$, where $m_2$ is a companion mass. This is
  because Gaia BH1, BH2, and NS1 have light companion stars: $0.93$,
  $1.07$, and $0.79 \msun$, respectively. This is also true for Gaia
  BH3 and a large fraction of Gaia NSs.
\item Located outside of open clusters. None of Gaia BHs and Gaia NSs
  are located in open clusters.
\end{enumerate}
Note that we do not put any constraints on orbital eccentricities in
the above criteria. On the other hand, Gaia BHs and NS1 are
characterized by moderately high eccentricities ($0.5$-$0.7$) and
small (but non-zero) eccentricity ($\sim 0.1$), respectively. Note
that the eccentricities of Gaia NSs range from $0.1$ to $0.8$. Compact
binaries we pick up may not be necessarily identified as Gaia BHs and
NSs. Thus, these compact binaries are hereafter called ``inert compact
binaries''. They are also called ``inert BH binaries (inert BHBs)''
and ``inert NS binaries (inert NSBs)'' if they contain BHs and NSs,
respectively.

\section{Results}
\label{sec:Results}

First, We make clear whether inert BHBs and NSBs are formed from
primordial binaries, or not. We find that no inert BHBs nor NSBs
originate from primordial binaries in all our models other than the
no-kick model. In the no-kick model, no inert BHBs originate from
primordial binaries, while inert NSBs can be formed from primordial
binaries and through dynamical interactions. Dynamical interactions
play an important role in forming inert BHBs regardless of the
presence of CCSN natal kicks. On the other hand, the formation of
inert NSBs strongly depend on CCSN natal kicks. In section
\ref{sec:StandardCcsnNatalKick}, we first investigate simulation
results of models with CCSN natal kicks. In section
\ref{sec:EffectsCcsnNatalKick}, we compare simulation results of
models with and without CCSN natal kicks.

\subsection{Standard CCSN natal kicks}
\label{sec:StandardCcsnNatalKick}

\begin{figure}
  \includegraphics[width=\columnwidth]{\fdir/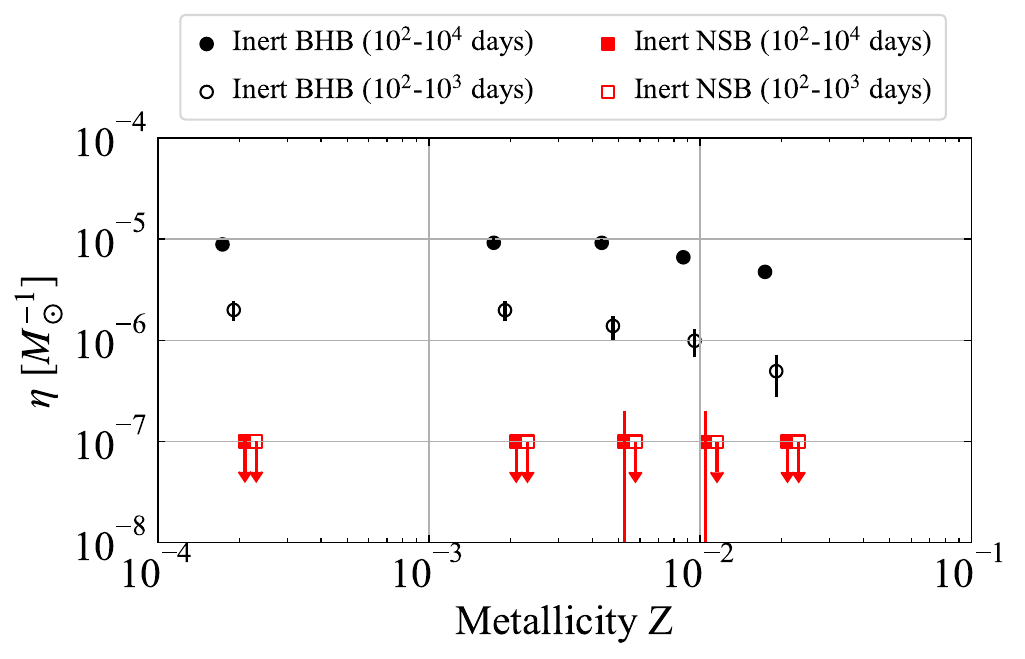}
  \includegraphics[width=\columnwidth]{\fdir/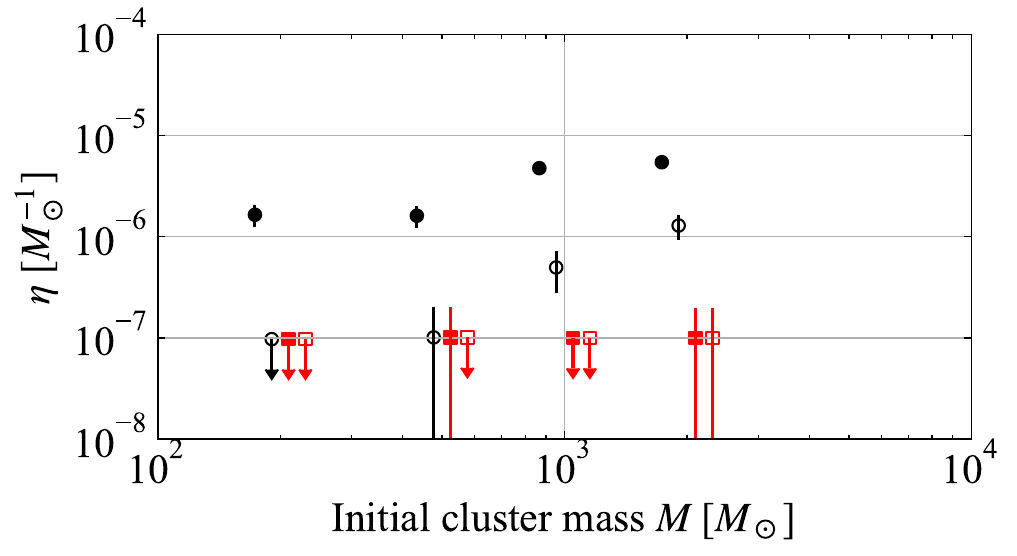}
  \includegraphics[width=\columnwidth]{\fdir/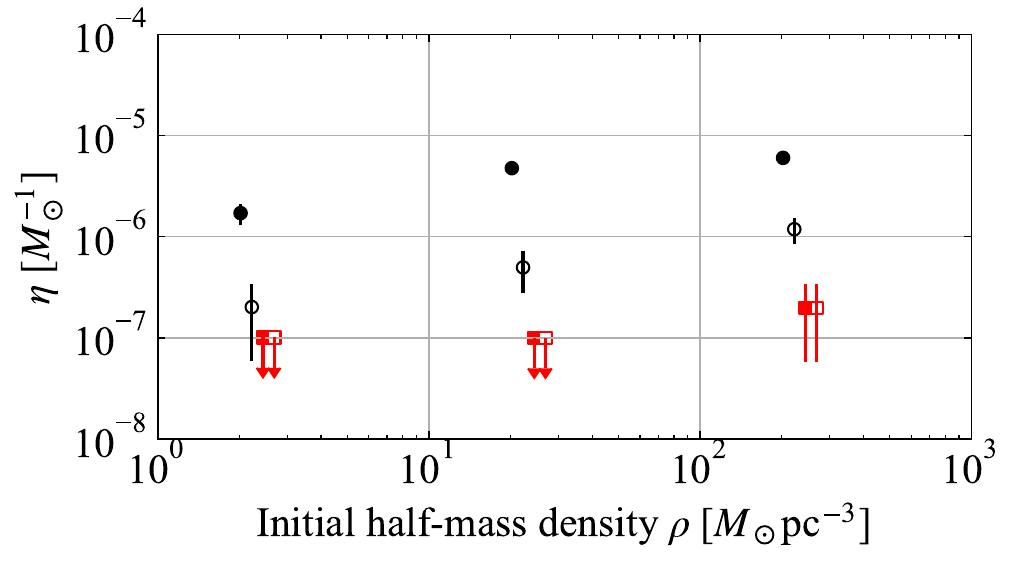}
  \includegraphics[width=\columnwidth]{\fdir/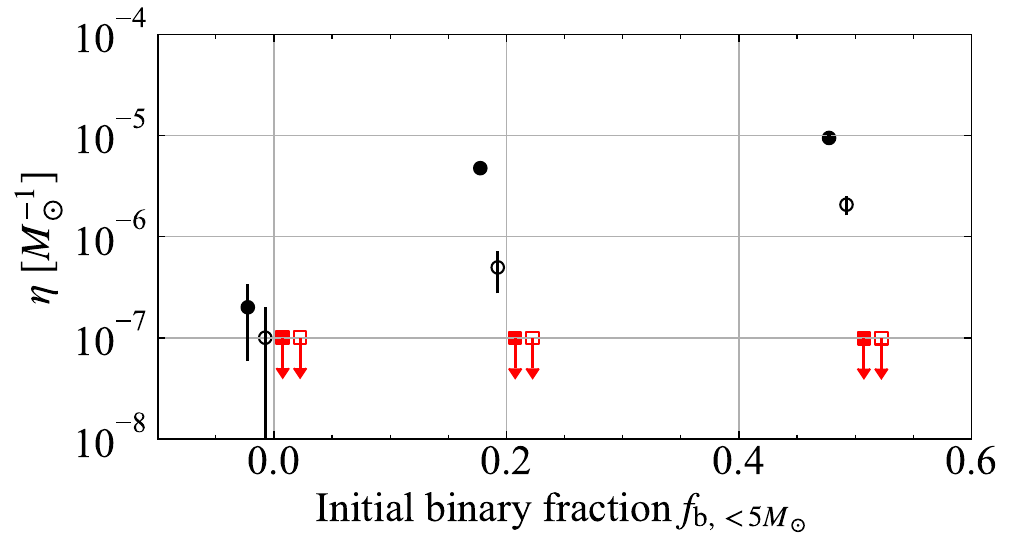}
  \caption{Formation efficiencies of inert BHBs (circles) and NSBs
    (squares) as a function of metallicity ($Z$), initial cluster mass
    ($M$), initial half-mass density ($\rho$), and initial binary
    fraction ($\fbl$). Filled and open points show that the ranges of
    binary orbital periods are $10^2$-$10^4$ days and $10^2$-$10^3$
    days, respectively. Circles and squares indicate inert BHBs and
    NSBs, respectively. These points are slightly shifted leftward or
    rightward for visibility. We calculate error bars, assuming that
    inert BHBs and NSBs are formed in Poisson process. Error bars in
    some models reach zero, since only one inert BHB or NSB is formed
    in these models. In several models, no inert NSBs are formed. For
    these models, we use down arrows to indicate the upper limits of
    inert NSB formation efficiencies.}
  \label{fig:gaiabhns}
\end{figure}

Figure \ref{fig:gaiabhns} shows the formation efficiencies of inert
BHBs and NSBs in all our cluster models other than the no-kick
model. Inert BHBs with $10^2 \le P/{\rm day} \le 10^4$ are formed at a
formation efficiency of $10^{-6}$-$10^{-5}\msun^{-1}$ except for the
$\fbl=0$ model. This is consistent with previous results
\citep{2023MNRAS.526..740R, 2024MNRAS.527.4031T,
  2024ApJ...965...22D}. The formation efficiencies slowly increase
with metallicity decreasing, initial cluster mass increasing, and
initial half-mass density increasing. The reason for the dependence on
initial cluster mass and half-mass density is that larger-mass, and
higher-density clusters are harder to be disrupted in the Galactic
tidal field, and thus these clusters have longer time to form inert
BHBs. We describe the reason for the dependence on metallicity
later. The formation efficiency in the $\fbl=0$ model is much smaller
than those in the other models for the following reason. Generally, an
inert BHB is formed, such that a single BH captures one star in a
binary. Since the number of binaries in the $\fbl=0$ model is much
smaller than those in the other models, inert BHBs are harder to form
in the $\fbl=0$ model. Nevertheless, it is worth remarking that inert
BHBs are formed at an efficiency of $\sim 10^{-7} \msun^{-1}$ even in
the $\fbl=0$ model. They would be formed through triple-single
encounters. This indicates that inert BHBs are robustly formed in open
clusters, even if an initial binary fraction is extremely low. This is
consistent with the results of \cite{2024MNRAS.527.4031T}.

Inert BHBs for $10^2 \le P/{\rm day} \le 10^3$ are formed at an
efficiency of $10^{-7}$-$10^{-6}\msun^{-1}$. In the $M=2000\msun$ and
$\rho=200$ models, the number ratio of such inert BHBs to all inert
BHBs is large, more than $0.1$. This would be because these clusters
have larger escape velocities than the other models. A cluster with a
larger escape velocity can retain closer binaries kicked by their
dynamical interactions, and so make them closer and closer. Although
the number ratio is also large in the $\fbl=0$ model, we cannot
identify if it is real, because the number of inert BHBs is too small.

The inert BHB formation efficiencies weakly increase with
metallicities decreasing. The formation efficiency in the $Z=0.0002$
model is 2 and 5 times than in the $Z=0.02$ model for inert BHBs with
$10^2 \le P/{\rm day} \le 10^4$ and $10^2 \le P/{\rm day} \le 10^3$,
respectively. We look into the reason of the dependence on
metallicity. This is because the dependence may be a key to elucidate
the origin of Gaia BHs as discussed in
\cite{2024arXiv240413047E}. Single BH formation efficiencies are not
sensitive to metallicities: $2.7 \times 10^{-3}$, $2.8 \times
10^{-3}$, $2.9 \times 10^{-3}$, $3.2 \times 10^{-3}$, and $3.0 \times
10^{-3} \msun^{-1}$ for $Z=0.02$, $0.01$, $0.005$, $0.002$, and
$0.0002$ models, respectively. Metallicities do not affect cluster
mass evolution: $\sim 700 \msun$, $\sim 200 \msun$ and $\lesssim 1
\msun$ at the clusters' ages of $40$, $400$, and $1000$ Myrs,
respectively, for all the metallicities. Thus, single BH formation
efficiency and cluster mass evolution are not the reason for the
dependence of the inert BHB formation efficiencies on
metallicities. On the other hand, the average single BH masses are
quite different among metallicities: $7.9$, $11.7$, $15.7$, $17.4$,
and $20.3 \msun$ for $Z=0.02$, $0.01$, $0.005$, $0.002$, and $0.0002$
models, respectively. Since more massive BHs can capture other stars
more easily, single BH masses should strongly affect the dependence of
the inert BHB formation efficiencies.

As seen in Figure \ref{fig:gaiabhns}, the formation efficiencies of
inert BHBs weakly increase with increasing cluster mass. On the other
hand, Figure 2 in \cite{2024arXiv240413036M} has shown that the
formation efficiencies decrease with increasing cluster mass. This
difference would come from two points. First, inert BHB formation is
inhibited due to more pronounced dynamical heating by larger BH
populations in more massive clusters, as pointed out by
\cite{2018ApJ...855L..15K} and \cite{2024arXiv240413036M}. The second
point is as follows. For clusters with $\lesssim 10^3 \msun$, two-body
relaxation (or thermal) timescale \citep{1987degc.book.....S} is
sufficiently shorter than the Hubble time. Thus, BHs can interact with
other stars for a long time with respect to clusters' thermal
timescales. However, their interactions are prevented by tidal
disruption of clusters. Since more massive clusters are more robust
against the Galactic tidal field, they can form inert BHBs more
efficiently. For clusters with $\gtrsim 10^5 \msun$, two-body
relaxation time is comparable to or longer than the Hubble time. From
the view point of clusters' thermal timescales, BHs have a shorter
time to interact with other stars in more massive clusters. Note that
the two-body relaxation time increases with increasing cluster
mass. In fact, the number of BHs decreases more slowly with increasing
cluster mass as seen in Figure 3 of \cite{2020ApJS..247...48K}, which
have published the CMC models used by \cite{2024arXiv240413036M}. This
clearly indicates that clusters' thermal ages are younger and BHs have
less chances to interact with other stars in more massive stars. About
$10^4 \msun$ clusters may form inert BHs most efficiently, because
they are hard to be tidally disrupted, and have a sufficiently shorter
two-body relaxation time than the Hubble time.

In contrast to inert BHBs, inert NSBs are hardly formed. The number of
inert NSBs is only 6 even if we put together all the models. It is
only 3 for inert NSBs with $10^2 \le P/{\rm day} \le 10^3$. If we
weigh equally all the models, the formation efficiency is $6 \times
10^{-8} \msun^{-1}$ for all the inert NSBs, and $3 \times 10^{-8}
\msun^{-1}$ for inert NSBs with $10^2 \le P/{\rm day} \le 10^3$. The
formation efficiency of inert NSBs is smaller than that of inert BHBs
by at least an order of magnitude.

\cite{2024OJAp....7E..27E, 2024arXiv240500089E} have not fully ruled
out the possibility that some of Gaia NSs are triple systems; for
example NSs are actually double WDs. This motivates us to investigate
triple and quadruple systems formed in all our cluster models. Such
systems include an inner MS binary with a single NS as well as an
inner double WDs with a single MS star. We do not find such a triple
system. Instead, we find only one triple system which is potentially
mistook for an inert NSB from the $Z=0.0002$ model. Its inner binary
consists of a $1.28 \msun$ NS and $0.022 \msun$ HeWD, and its period
and eccentricity are $32$ minutes and $0.023$, respectively. This
inner binary is formed from a primordial binary. The outer binary has
the period and eccentricity of $560$ days and $0.76$, respectively,
and the other outer component is a $0.95 \msun$ MS. The inner binary
is in the process of mass transfer from the HeWD to the NS. This mass
transfer is stable and long-lasting. This is the reason why the inner
binary can capture the third star. Because of the mass transfer, the
triple system should be observed as an X-ray binary, not as an inert
NSB. Thus, we exclude the triple system from our list of inert
NSBs. Even if we add it to our list, the formation efficiency of inert
NSBs is still much smaller than that of inert BHBs.

\begin{figure}
  \includegraphics[width=\columnwidth]{\fdir/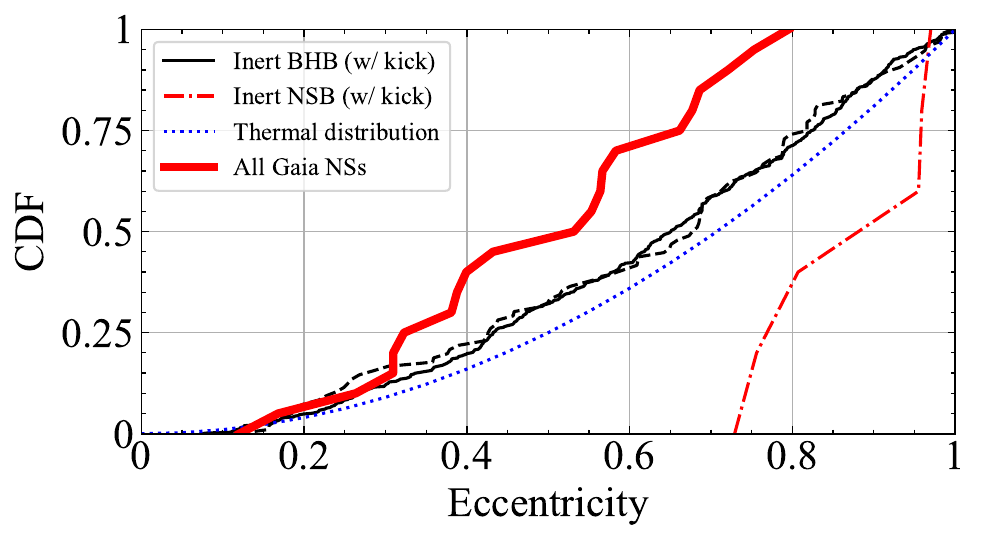}
  \caption{Cumulative eccentricity distribution of inert BHBs and
    NSBs. Solid and dashed curves indicate inert BHBs with binary
    orbital periods of $10^2$-$10^4$ and $10^2$-$10^3$ days,
    respectively. The dot dashed curve indicates inert NSBs in all the
    models. The thick solid curve indicates all the Gaia NSs.
    The dotted curve shows the thermal distribution.}
  \label{fig:gaiabhnsEccentricity}
\end{figure}

Figure \ref{fig:gaiabhnsEccentricity} shows the orbital eccentricity
distribution of inert BHBs and NSBs. We put together inert BHBs and
NSBs among all the models other than the no-kick model. We have to put
together inert NSBs, because the number of inert NSBs is too small to
statistically investigate the eccentricity distribution of inert NSBs
for each model. We adopt the Kolmogorov-Smirnov (K-S) test to compare
these distribution with the thermal distribution
\citep{1975MNRAS.173..729H} and Gaia BHs and NSs. Note that Gaia BH1,
BH2 and NS1 have orbital eccentricities of $\sim 0.451$
\citep{2023MNRAS.518.1057E}, $0.518$ \citep{2023MNRAS.521.4323E}, and
$0.124$ \citep{2024OJAp....7E..27E}, respectively. After this paper
was submitted, \cite{2024arXiv240500089E} have also published the
orbital parameters of all Gaia NSs. We compare inert NSBs not only
with Gaia NS1, but also with all the Gaia NSs. Table
\ref{tab:PvalueKstest} shows the P-values of the K-S test. The
eccentricity distribution of inert BHBs with $10^2 \le P/{\rm day} \le
10^4$ is clearly more circular than the thermal distribution, since
the P-value is $5.6 \times 10^{-5}$. This is possibly because
eccentric BHBs are circularized by tidal interactions. This is also
seen in the eccentricity-period diagram of binaries in Gaia DR3
\citep[e.g.][]{2023A&A...674A..34G}. On the other hand, the
eccentricity distribution of inert BHBs with $10^2 \le P/{\rm day} \le
10^3$ is not significantly different from the thermal distribution;
the P-value is $9.3 \times 10^{-2}$. This may be due to their small
number.

\begin{table}
  \centering
  \caption{P-values of the Kolmogorov-Smirnov (K-S) test for
    eccentricity distributions of inert BHBs and
    NSBs.} \label{tab:PvalueKstest}
  \begin{tabular}{lccc}
    \hline
    Binary type & Number & Comparison & P-value \\
    \hline
    Inert BHB ($10^2$-$10^4$ day) & 556 & Thermal  & $5.6 \times 10^{-5}$ \\
    Inert BHB ($10^2$-$10^3$ day) &  97 & Thermal  & $9.3 \times 10^{-2}$ \\
    Inert NSB ($10^2$-$10^4$ day) &   6 & Thermal  & $4.2 \times 10^{-2}$ \\
    Inert NSB selected            &   4 & Thermal  & $1.4 \times 10^{-1}$ \\
    Inert NSB (no-kick)           &  34 & Thermal  & $2.5 \times 10^{-3}$ \\
    \hline
    Inert BHB ($10^2$-$10^4$ day) & 556 & Gaia BHs & $6.4 \times 10^{-1}$ \\
    Inert BHB ($10^2$-$10^3$ day) &  97 & Gaia BHs & $6.9 \times 10^{-1}$ \\
    Inert NSB ($10^2$-$10^4$ day) &   6 & Gaia NS1 & $2.9 \times 10^{-1}$ \\
    Inert NSB selected            &   4 & Gaia NS1 & $4.0 \times 10^{-1}$ \\
    Inert NSB (no-kick)           &  34 & Gaia NS1 & $5.7 \times 10^{-2}$ \\
    Inert NSB ($10^2$-$10^4$ day) &   6 & All Gaia NSs & $1.9 \times 10^{-4}$ \\
    Inert NSB selected            &   4 & All Gaia NSs & $2.4 \times 10^{-3}$ \\
    Inert NSB (no-kick)           &  34 & All Gaia NSs & $1.0 \times 10^{-5}$ \\
    \hline
  \end{tabular}
\end{table}

Although the eccentricity distribution of inert NSBs appears to be
largely different from the thermal distribution, it is still
consistent with the thermal distribution. The P-value is not small:
$4.2 \times 10^{-2}$. This may be again due to the small
statistics. As described later, the inert NSBs with the highest and
second highest eccentricities are not formed through dynamical
interactions (named ``CCSN channel'' later), although they are not
primordial binaries. Moreover, these inert NSBs are formed only in the
$\rho=200$ model. A large fraction of open clusters in reality could
be less dense than the $\rho=200$ model. Thus, we construct inert
NSBs, excluding these NSBs (``inert NSB selected'' in Table
\ref{tab:PvalueKstest}). When we compare their eccentricity
distribution with the thermal distribution, the P-value increases to
$1.4 \times 10^{-1}$. It is natural that they are formed through
dynamical interactions. Nevertheless, their eccentricity distribution
may finally becomes more circular than the thermal distribution,
similarly to inert BHBs, if we obtain more many inert NSBs with
simulating more many clusters.

Inert BHBs and NSBs have the eccentricity distributions which are not
inconsistent with Gaia BHs and Gaia NS1, respectively, regardless of
how to choose these BHBs and NSBs. The P-values are more than
$0.1$. This may be due to the small numbers; the number of inert NSBs
are only 6, and the numbers of Gaia BHs and Gaia NS1 are 2 and 1,
respectively. In fact, when we compare our inert NSBs with all the
Gaia NSs, the P-values are quite small: $1.9 \times 10^{-4}$ for all
the inert NSBs, and $2.4 \times 10^{-3}$ for the inert NSBs
selected. Owing to the publication of Gaia NSs' eccentricities, we can
find that the eccentricity distribution of our inert NSBs is different
from that of Gaia NSs. On the other hand, It seems that the
eccentricity distribution of Gaia BHs is not currently helpful to
assess if inert BHBs are identified as Gaia BHs.

\begin{figure}
  \includegraphics[width=\columnwidth]{\fdir/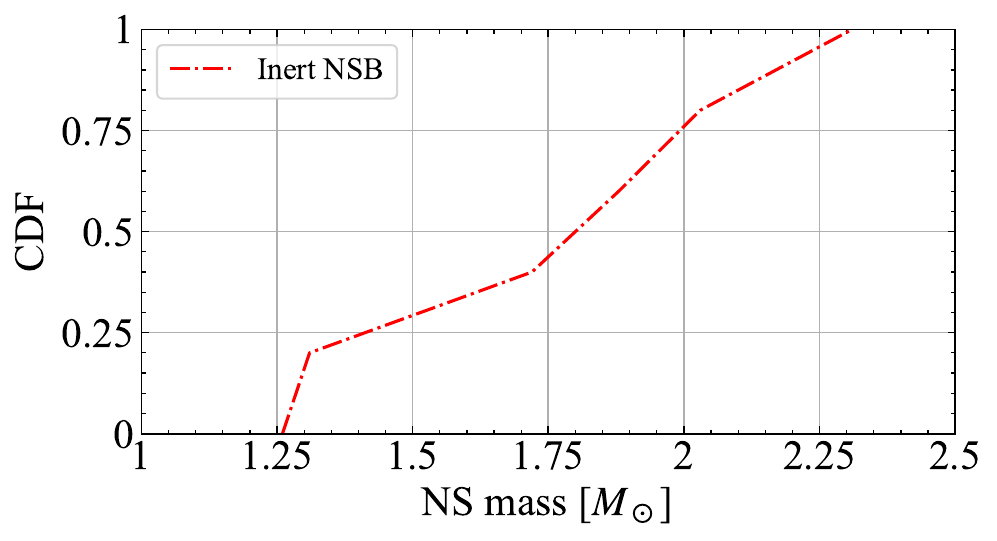}
  \caption{Cumulative NS mass distribution of inert NSBs in all the
    models.}
  \label{fig:gaiabhnsPrimarymass}
\end{figure}

Figure \ref{fig:gaiabhnsPrimarymass} shows the NS mass distribution of
all the inert NSBs. 2 of the 6 inert NSBs have NSs with $>2
\msun$. These NSs are formed through AIC of WD merger products (called
``AIC channel'' later). The detail formation channel is described
below.

\begin{figure}
  \includegraphics[width=\columnwidth]{\fdir/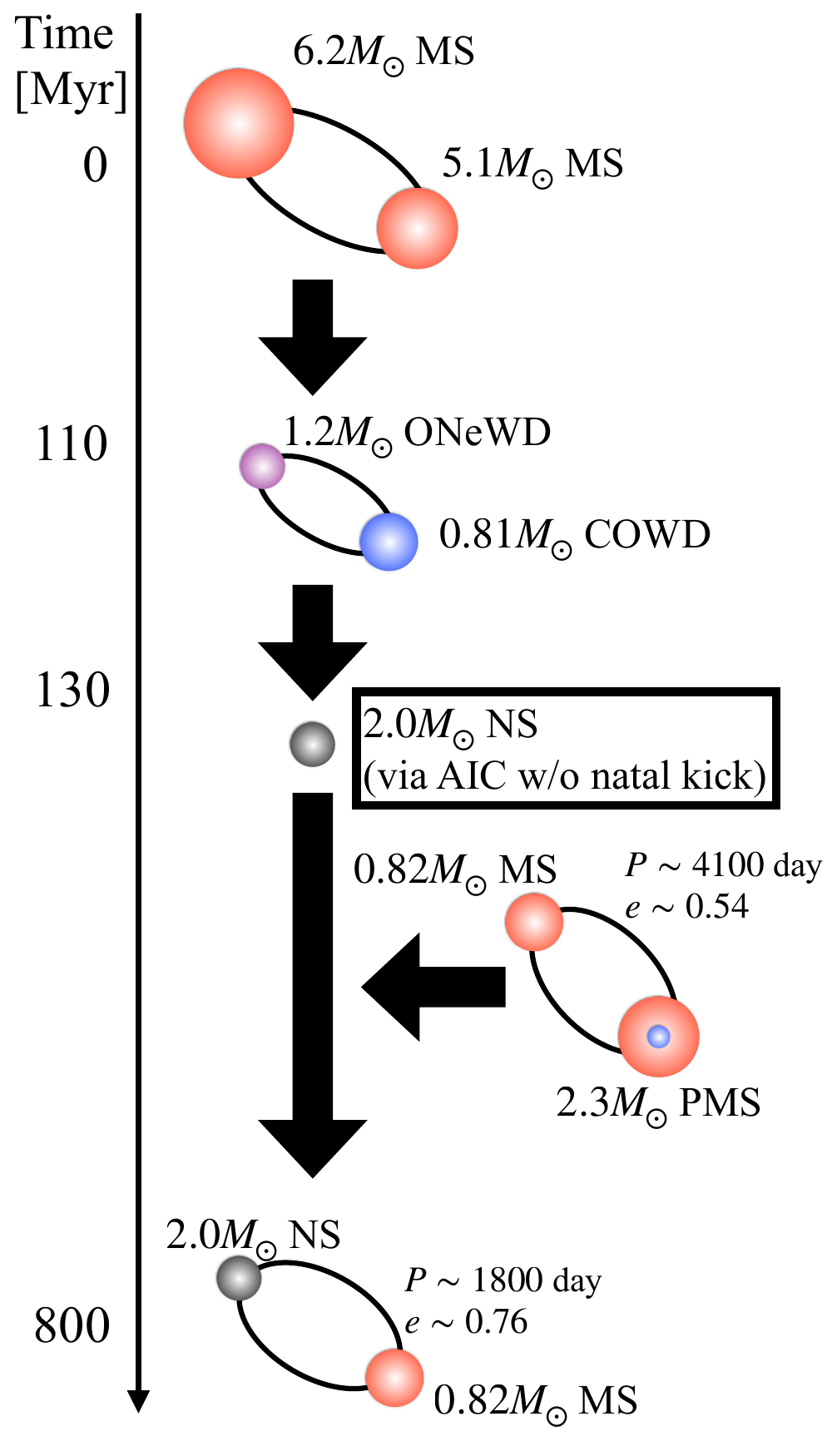}
  \caption{Inert NSB formation channel via AIC. This happens in model
    $Z=0.005$. A binary with $6.2$ and $5.1 \msun$ MSs evolves to
    binary WDs with $1.2 \msun$ ONeWD and $0.81 \msun$ COWD. These WDs
    merge, and its merger product collapses to a NS. Because this NS
    receives no natal kick, it stays inside of the open cluster. The
    NS closely interacts with a binary with $0.82 \msun$ MS and $2.3
    \msun$ PMS, and capture the $0.82 \msun$ MS, resulting in an inert
    NSB. It finally escapes from the open cluster.}
  \label{fig:gaians_aic}
\end{figure}

\begin{figure}
  \includegraphics[width=\columnwidth]{\fdir/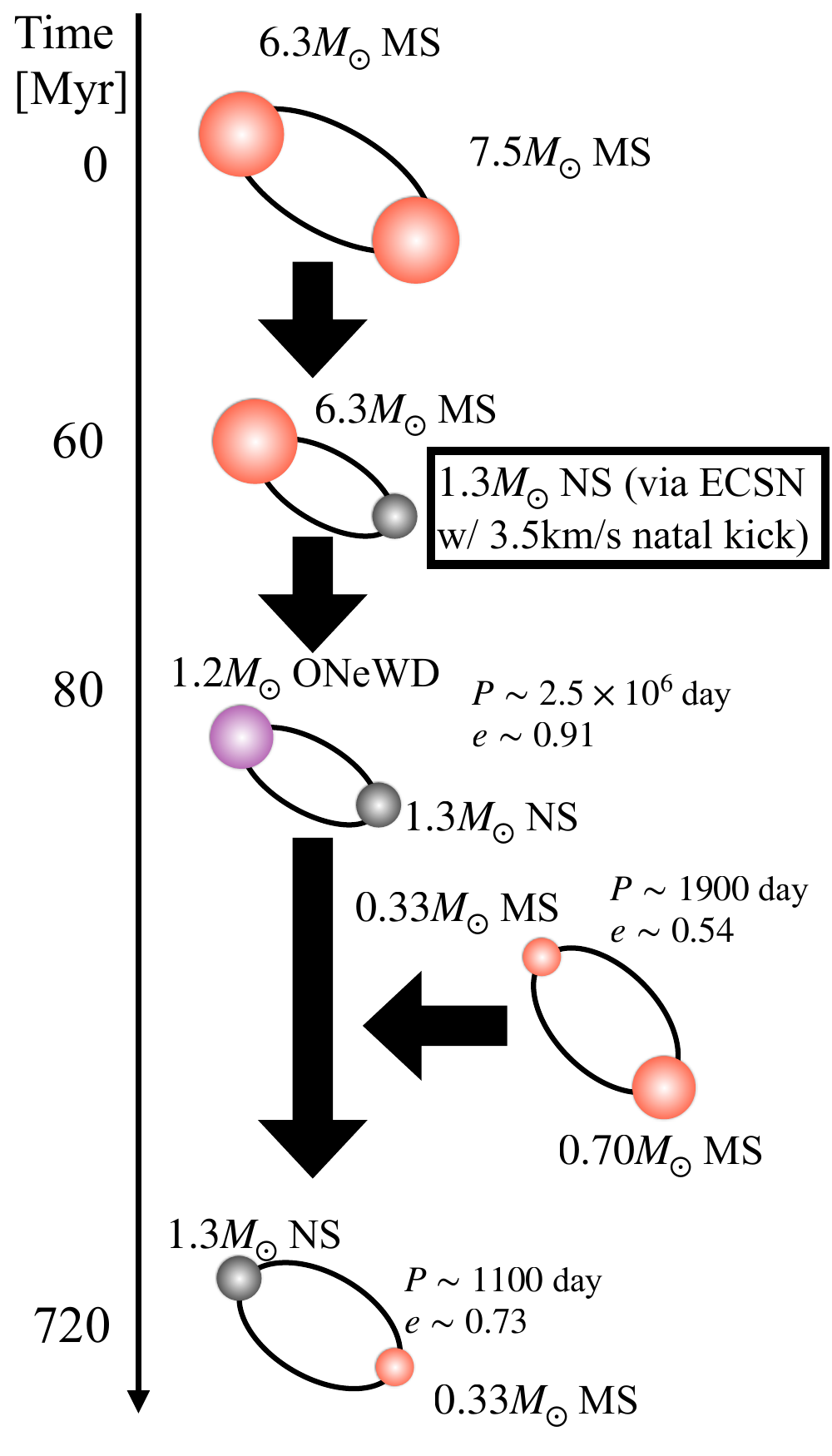}
  \caption{Inert NSB formation channel via ECSN. This happens in model
    $Z=0.01$. A binary with $7.5$ and $6.3 \msun$ MSs evolves to a
    binary with $1.3 \msun$ NS and $1.2 \msun$ ONeWD. The NS receives
    a small kick ($3.5$ km~s$^{-1}$) because of ECSN, and is retained
    in the open cluster. The binary interacts with a binary with
    $0.70$ and $0.33 \msun$ MSs, and leaves an inert NSB consisting of
    the NS and $0.33 \msun$ MS. The inert NSB finally escapes from the
    open cluster.}
  \label{fig:gaians_ecsn}
\end{figure}

\begin{figure}
  \includegraphics[width=\columnwidth]{\fdir/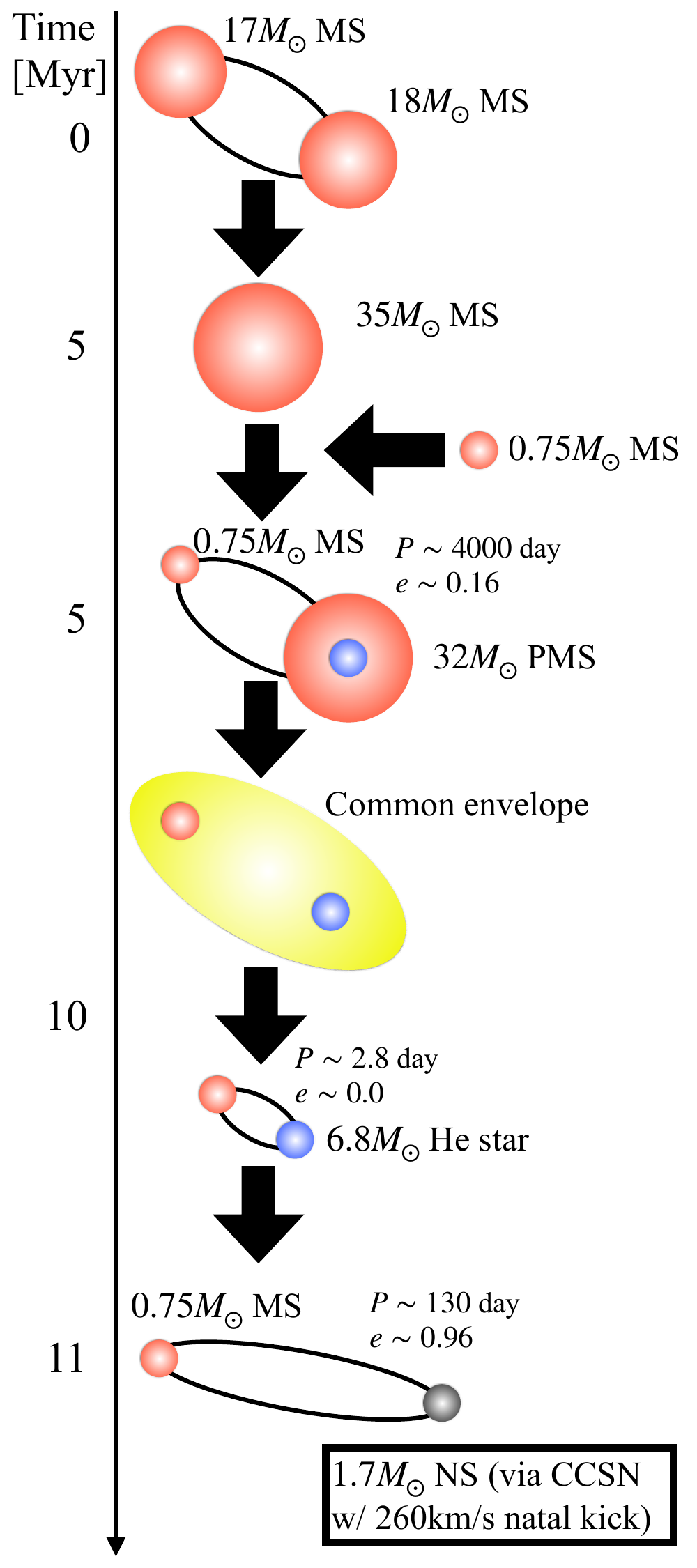}
  \caption{Inert NSB formation channel via CCSN. This happens in model
    $\rho=200$. A binary with $18$ and $17 \msun$ MSs merges to a $35
    \msun$ MS. The $35 \msun$ MS soon captures a $0.75 \msun$ MS. This
    binary experiences common envelope evolution, and results in a
    close binary with $6.8 \msun$ He star and $0.75 \msun$ MS. The He
    star undergoes CCSN, and creates a $1.7 \msun$ NS. Although the NS
    receives a large kick ($260$ km~s$^{-1}$), the binary is not
    disrupted because of the close separation, and becomes an inert
    NSB. On the other hand, the large kick ejects the inert NSB from
    the open cluster.}
  \label{fig:gaians_ccsn}
\end{figure}

Hereafter, we describe three formation channels of inert NSBs. These
channels are illustrated in Figures \ref{fig:gaians_aic},
\ref{fig:gaians_ecsn}, and \ref{fig:gaians_ccsn}. In the first
channel, NSs are formed through AIC of WD merger products. We call
this channel ``AIC channel''. We explain the AIC channel along with
Figure \ref{fig:gaians_aic}. This channel starts with a primordial
binary consisting of intermediate mass stars: $6.2$ and $5.1 \msun$
MSs. They evolve to a close double WDs with $1.2 \msun$ ONeWD and
$0.81 \msun$ COWD via common envelope evolution. They merge through
orbital decay due to gravitational wave radiation. The merger product
avoids type Ia SN explosion, and collapses to a NS (AIC). The NS mass
is $> 2 \msun$. The NS can avoid escaping from the open cluster, since
AIC generate no natal kick in our model. Finally, the NS dynamically
interacts with another primordial binary, and captures one star of the
primordial binary. The resulting NSB has orbital period and
eccentricity of $\sim 1800$ day and $\sim 0.76$, respectively. This
happens in the $Z=0.005$ model. Another one also occurs in the
$M=2000\msun$ model.

Our simulations implement a NS formation channel through AIC of WDs
accreting from non-degenerate companions. In fact, we find such NS
formations. However, they do not evolve to inert NSBs. Their orbital
periods are much shorter than $10^2$ days, because such NS formations
involve mass transfer from non-degenerate companions. These binaries
are too close to interact with other stars dynamically. Moreover, if
they experience dynamical interactions, their orbital periods will be
still much shorter than than $10^2$ days. Such tight binaries tend to
become tighter through dynamical interactions.

The second channel involves ECSNe as seen in Figure
\ref{fig:gaians_ecsn}. We call this channel ``ECSN channel''. We find
this channel in the $Z=0.01$ and $M=500\msun$ models. Here, we
introduce the formation processes of the inert NSB in the former model
(see also Figure \ref{fig:gaians_ecsn}). A $7.5\msun$ MS with a $6.3
\msun$ MS companion evolves to a $1.3 \msun$ NS via ECSN. Because of
ECSN, the NS receives a small kick velocity ($3.5$ km~s$^{-1}$), and
is not ejected from the open cluster. Its companion evolves to a $1.2
\msun$ ONeWD. This binary experiences no interaction any more, since
their separation is large. Very later, this binary interacts with
another binary with $0.33$ and $0.70 \msun$ MSs. This interaction
raises an inert NSB with the $1.3 \msun$ NS and $0.33 \msun$ MS. Note
that, although a lower-mass star is harder to be took as a binary
member, it is not impossible.  This has orbital period and
eccentricity of $\sim 1100$ day and $0.73$, respectively.

In the third channel, NSs arise from CCSNe. Thus, we refer to them as
``CCSN channel''. We obtain two inert NSBs, both of which are formed
in the $\rho=200$ model. In Figure \ref{fig:gaians_ccsn}, we show the
formation pathway of one of the two. This channel begins with a
primordial binary with $18$ and $17\msun$ MSs. The primordial binary
merges to a $35 \msun$ MS. Just after the merger, the $35 \msun$ MS
soon captures a $0.75\msun$ MS. It evolves to a $32 \msun$ PMS, and
experiences common envelope evolution with its $0.75\msun$ MS
companion. The common envelope evolution leaves a short-period ($2.8$
day) binary with a $6.8 \msun$ He star and $0.75 \msun$ MS. The He
star causes a CCSN with a kick velocity of $260$ km~s$^{-1}$. It
luckily survives as a binary. Such a close binary can survive a CCSN
depending on the kick velocity direction despite of CCSN mass loss,
because its binary internal motion has large velocity comparable to
the kick velocity. However, the natal kick greatly affects the
center-of-mass velocity of the binary, and ejects the resulting NSB
from the open cluster.

The CCSN channel involves a dynamical interaction to form the
progenitors of inert NSBs, while the final configurations of inert
NSBs are dependent on binary physics, not on dynamical
interactions. Furthermore, this channel tends to inert NSBs with high
eccentricities. In fact, the two NSBs formed through the CCSN channel
have the highest and second highest eccentricities among inert NSBs in
all our models other than the no-kick model. Thus, we exclude inert
NSBs via the CCSN channel from the ``inert NSBs selected'' in Table
\ref{tab:PvalueKstest} when we compare their eccentricity distribution
with the thermal distribution.

The CCSN channel appears only in the $\rho=200$ model. We interpret
the reason as follows. This channel requires a post common envelope
binary consisting of $<= 1.1 \msun$ MS star and massive He star (say
$\gtrsim 3 \msun$). If a He star has $\lesssim 3 \msun$, such a post
common envelope binary cannot become wide enough to be an inert NSB
after the He star collapses to a NS. This is because the He star does
not lose its mass during its CCSN. The kick velocity is at most
comparable to the velocity of the binary internal motion, and thus
cannot change the binary period largely. A He star with $\gtrsim 3
\msun$ is raised by a $\gtrsim 11 \msun$ MS. However, our initial
conditions prohibit primordial binaries with $\gtrsim 11 \msun$ and
$\lesssim 1.1\msun$ MS stars, since the minimum binary mass ratio is
$0.1$. Thus, binaries with $\gtrsim 11 \msun$ and $\lesssim 1.1\msun$
MS stars need to be formed through dynamical interactions before
$\gtrsim 11 \msun$ stars evolve to NSs or BHs. Dynamical interactions
in the $\rho=200$ model are the most active among all our models
because of its high density. Thus, many binaries with $\gtrsim 11
\msun$ and $\lesssim 1.1\msun$ MS stars are formed, and there are many
chances to achieve the CCSN channel in the $\rho=200$ model. We have
to note that this reason strongly depends on our initial conditions in
which there are no primordial binaries with $\gtrsim 11 \msun$ and
$\lesssim 1.1\msun$ MS stars. It would not be strange if The CCSN
channel is achieved in real open clusters with the initial half-mass
density of $< 200 \msun~{\rm pc}^{-3}$.

If we reduce the minimum binary mass ratio, another inert NSB
formation channel (hereafter called primordial binary or PB channel)
similar to the CCSN channel may appear in primordial binaries, and in
isolated binaries. The difference between the two channels is how to
form progenitors of inert NSBs. The progenitors are formed through
dynamical interactions in the CCSN channel, while they are present
from the initial time in the PB channel. The PB channel increases the
formation efficiency of inert NSBs (possibly BHBs) not only in open
clusters, but also on isolated fields. In this case, the importance of
open clusters would diminish. This is because the number of channels
increases from 3 to 4 in open clusters, while it increases from 0 to 1
on isolated fields.

\subsection{Effects of CCSN natal kicks}
\label{sec:EffectsCcsnNatalKick}

\begin{figure}
  \includegraphics[width=\columnwidth]{\fdir/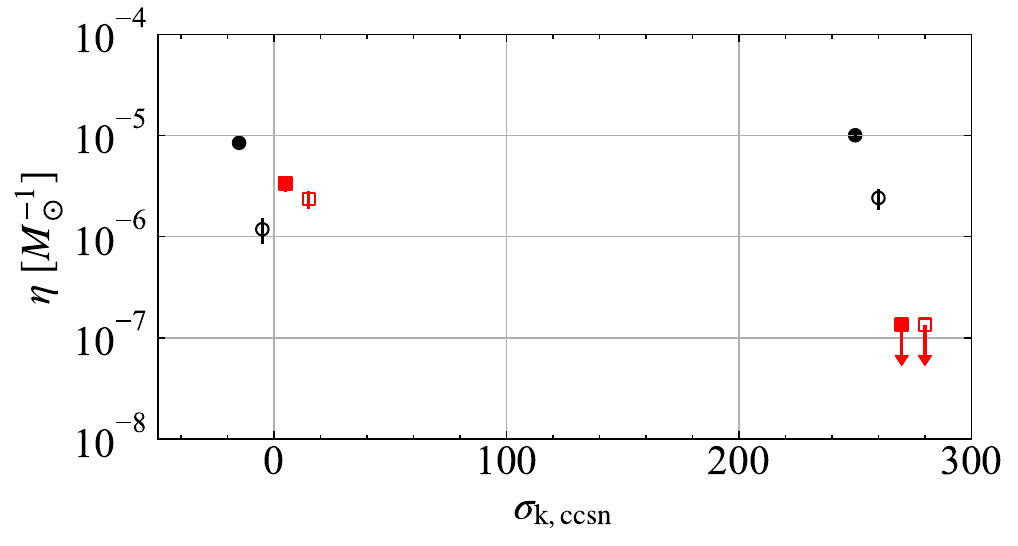}
  \caption{The same as Figure \ref{fig:gaiabhns} except for a function
    of CCSN natal kick velocity ($\sigmakccsn$). The results of
    $Z=0.002$ and no-kick models are shown.}
  \label{fig:gaiabhns_nokick}
\end{figure}

Hereafter, we investigate the effects of CCSN natal kicks, showing the
simulation results of the no-kick model. Figure
\ref{fig:gaiabhns_nokick} shows the formation efficiency of inert BHBs
and NSBs in the $Z=0.002$ and no-kick models. The formation efficiency
of inert BHBs in the no-kick model is similar to that in the other
models regardless of $10^2 \le P/{\rm day} \le 10^4$ and $10^2 \le
P/{\rm day} \le 10^3$. On the other hand, NSBs are formed much more
efficiently in the no-kick model (a few $10^{-6} \msun^{-1}$) than in
the other models ($\lesssim 10^{-7} \msun^{-1}$). There are $34$ inert
NSBs with $10^2 \le P/{\rm day} \le 10^4$. $25$ of them originate from
primordial binaries, and the rest of them are formed through dynamical
interactions.

The no-kick model has a high formation efficiency of inert NSBs, in
contrast to other models. As described above, a dominant fraction of
inert NSBs ($\sim 70$ \%) originate from primordial binaries. We
describe their formation pathway here. We do not illustrate it,
because we suppose the no-kick model is unrealistic. Let's prepare a
binary with $\sim 8\msun$ and $\sim 1 \msun$ MSs. It separation is
$\sim 3 \times 10^3 \rsun$. The heavier star evolves to a PMS, and
fills its Roche lobe. Then, the binary experiences common envelope
evolution. The common envelope evolution leaves a binary with $\sim 2
\msun$ He star and $\sim 1 \msun$ MS star. It is separated by $\sim 40
\rsun$. The He star causes a CCSN, and leaves behind a NS. Because of
CCSN mass loss, the binary is widened, and fits in an inert
NSB. In all the models with CCSN natal kicks, such inert NSBs cannot
be formed. CCSN natal kicks disrupt the binary at a high
probability. Just before CCSNe, the velocities of binary internal
motions have $\sim 100$ km~s$^{-1}$, much smaller than CCSN natal kick
velocities ($265$ km~s$^{-1}$).

$9$ of $34$ inert NSBs are dynamically formed. For $6$ of them, NSs
capture their companions. On the other hand, for the rest of them, NS
progenitors capture their companions, and later they collapse to
NSs. The latter formation pathway is the same as the formation pathway
from primordial binaries descried above, except that a binary with
$\sim 8\msun$ and $\sim 1 \msun$ MSs is formed through dynamical
interactions.

\begin{figure}
  \includegraphics[width=\columnwidth]{\fdir/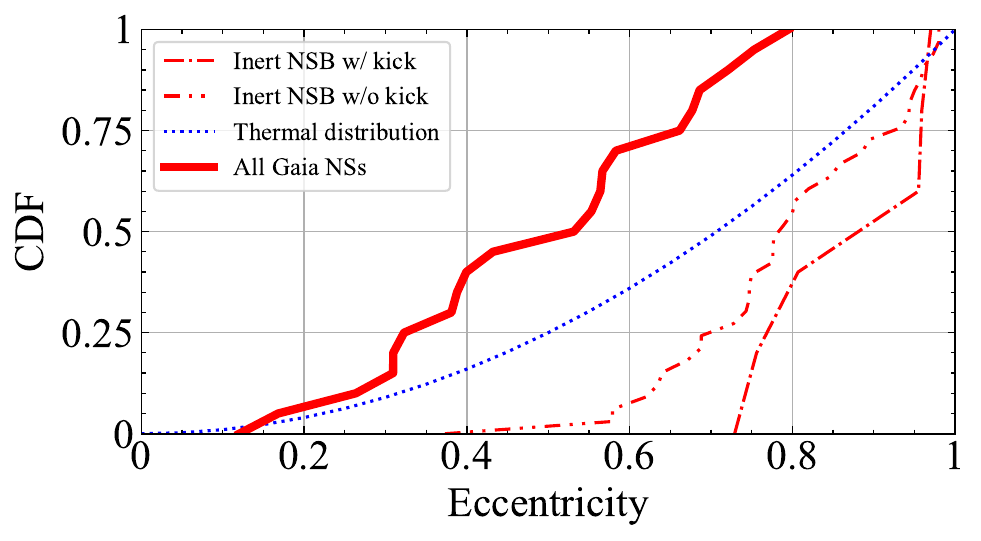}
  \caption{Cumulative eccentricity distribution of inert NSBs in all
    models except for the no-kick model (Inert NSB w/ kick), and in
    the no-kick model (Inert NSB w/o kick). The thick solid curve
      indicates all the Gaia NSs. The dotted curve shows the thermal
    distribution.}
  \label{fig:gaianokickEccentricity}
\end{figure}

Figure \ref{fig:gaianokickEccentricity} shows the cumulative
eccentricity distribution of inert NSBs in the no-kick model. Their
orbital eccentricities appear to deviate from the thermal
distribution. In fact, the deviation is significant; the P-value of
the K-S test is $2.5 \times 10^{-3}$ as seen in Table
\ref{tab:PvalueKstest}. This is because a dominant fraction of these
inert NSBs originate from primordial binaries, not formed through
dynamical interactions. On the other hand, the eccentricity
distribution of these NSBs is not significantly different from Gaia
NS1; the P-value is $5.7 \times 10^{-2}$.  This shows that it is
difficult to distinguish the differences among these eccentricity
distributions due to the small statistics if we use only Gaia
NS1. However, when we use all the Gaia NSs, the P-value is small, $1.0
\times 10^{-5}$. Because of improved statistics, we can conclude that
the eccentricity distribution of inert NSBs in the no-kick model is
different from that of all Gaia NSs.

When we switch off CCSN natal kicks, the number of NSs retained in
open clusters should increase by about 100 times. Nevertheless, the
number of inert NSBs dynamically formed is still low for the following
reason.  NSs are hard to capture other stars dynamically, because
their masses are comparable to or less than surrounding main-sequence
(MS) stars. Note that the lifetime of a $1.4 \msun$ MS star is a few
Gyr, which is longer than the lifetime of open clusters.

\section{Discussion}
\label{sec:Discussion}

First, we discuss whether inert BHBs and NSBs correspond to Gaia BHs
and NSs, respectively. Here, we do not consider the simulation results
of the no-kick model. As seen in Table \ref{tab:PvalueKstest}, the
P-values of the K-S test do not show that inert BHBs have different
orbital eccentricities from Gaia BHs. The eccentricity distributions
of inert NSBs with $10^2 \le P/{\rm day} \le 10^4$ and inert NSBs
selected in Table \ref{tab:PvalueKstest} are not inconsistent with
Gaia NS1's eccentricity. However, when we consider all the Gaia NSs,
the eccentricity distribution of inert NSBs is different from that of
all Gaia NSs. Thus, it is difficult to identify our inert NSBs as Gaia
NSs.

Even if we identify our inert NSBs as Gaia NSs, there is another
problem. We compare the formation efficiencies of inert BHBs and NSBs,
assuming that inert NSBs correspond to Gaia NSs. The formation
efficiency of inert NSBs is smaller than that of inert BHBs by two or
more orders of magnitude except for the no-kick model (see Figure
\ref{fig:gaiabhns}). However, the number of discovered Gaia NSs, $21$
\citep{2024OJAp....7E..27E, 2024arXiv240500089E}, is much larger than
the number of Gaia BHs, $2$. This shows that open clusters cannot form
enough Gaia NSs to be already discovered.

This is also true for the no-kick model. As for inert BHBs and NSBs
with $10^2 \le P/{\rm day} \le 10^4$, the formation efficiency of
inert NSBs is smaller than that of inert BHBs. Actually, the formation
efficiency of inert NSBs with $10^2 \le P/{\rm day} \le 10^3$ is
larger than that of inert BHBs with $10^2 \le P/{\rm day} \le
10^3$. However, a dominant fraction of inert NSB formed in the no-kick
model ($\sim 74$ \%) originate from primordial binaries. This means
that such inert NSBs can be formed through isolated binary evolution
if CCSN natal kicks are smaller than expected. In this case, although
a large number of inert NSBs can be formed in open clusters, inert
NSBs are also formed more efficiently on isolated fields. Thus, open
clusters cannot dominate the formation of Gaia NSs, even if CCSN natal
kicks are unrealistically small.

It is unclear how efficiently Gaia \citep{2023A&A...674A..34G} and its
spectroscopic follow-up observations \citep{2023MNRAS.518.1057E,
  2023MNRAS.521.4323E, 2024OJAp....7E..27E, 2024arXiv240500089E,
  2023AJ....166....6C, 2024arXiv240410486G} can discover Gaia BHs and
NSs. Thus, the numbers of discovered Gaia BHs and NSs do not directly
mean their intrinsic numbers. However, for the following reason, we
infer that Gaia NSs are harder to be discovered than Gaia BHs, and
that the intrinsic number of Gaia NSs should be much larger than that
of Gaia BHs. All Gaia and its follow-up observations search for Gaia
BHs and NSs, observing BHs' and NSs' companions. Because such
companions are swayed more largely with BH and NS masses increasing,
Gaia, astrometric observations, can more easily discover BHs and NSs
with larger masses. Similarly, companions tend to have larger radial
velocity variations, and thus spectroscopic follow-up observations can
detect more massive BHs and NSs more efficiently. Since NSs are
generally less massive than BHs, the intrinsic number ratio of Gaia
NSs to Gaia BHs should be larger than the reported number ratio. As
seen in section \ref{sec:Results}, open clusters form Gaia BHs more
efficiently than Gaia NSs. This is in contrast to the observational
results. Eventually, open clusters cannot form both of Gaia BHs and
Gaia NSs.

As described above, Gaia NSs are unlikely to be formed in open
clusters. Nevertheless, we do not mean that Gaia BHs are not formed in
open clusters. The formation efficiency of Gaia BHs in $\sim 10^3
\msun$ open clusters is still high ($\sim 10^{-6}$-$10^{-5} \msun$),
if we do not adopt unrealistically small half-mass density ($\rho \sim
2$ $\msun~{\rm pc}^{-3}$) nor unrealistically small binary fraction
($\fbl \sim 0$). It is reasonable that open clusters have $\rho
\gtrsim 20$ $\msun~{\rm pc}^{-3}$ initially. Note that, although
nearby open clusters have $\rho \sim 10^{-2}$-$10^2$ $\msun~{\rm
  pc}^{-3}$, they should become less dense with time through gas
expulsion, stellar evolution mass loss, and two-body relaxation
\citep{2010ARA&A..48..431P}. Initial binary fractions in open clusters
should be $\fbl \gtrsim 0.2$, similarly to isolated fields
\citep{2012Sci...337..444S, 2017ApJS..230...15M}. It would be possible
that Gaia BHs are formed in open clusters, while Gaia NSs
\citep{2024OJAp....7E..27E} and ultramassive WDs
\citep{2024MNRAS.527.11719} are formed on isolated fields. This will
become clear after much more Gaia BHs will be discovered.

\section{Summary}
\label{sec:Summary}

We have performed gravitational $N$-body simulations for $1.3 \times
10^8 \msun$ open clusters in total with various masses, metallicities,
densities, binary fractions, and natal kick velocities. We have found
that the formation efficiency is $10^{-6}$-$10^{-5} \msun^{-1}$ for
Gaia BHs in open clusters with realistic parameters, while $\lesssim
10^{-7} \msun^{-1}$ for Gaia NSs. This is in contrast to observational
results in which the number of discovered Gaia BHs ($3$) is smaller
than that of Gaia NSs ($21$). Moreover, inert NSBs obtained from our
simulations have different orbital eccentricities from Gaia NSs. Note
that inert BHBs obtained from our simulations have orbital
eccentricities not inconsistent with Gaia BHs. This is partly because
of the small statistics for Gaia BHs.

We have switched off CCSN natal kicks in order to increase the number
of NSs retained in open clusters. Although we expected the number of
inert NSBs to increase, the formation efficiency of inert NSBs is at
most comparable to that of inert BHBs. NSs are much harder to capture
companions than BHs, since NSs are not so heavier than MSs and WDs
during open cluster lifetimes ($\sim 1$ Gyr). Moreover, a dominant
fraction of them are formed from primordial binaries. Thus, inert NSBs
are less efficiently formed in open clusters than on isolated fields,
even if CCSNe generate no natal kicks.

Regardless of the presence and absence of CCSN natal kicks, inert NSBs
are formed in open clusters less efficiently than, or at most
comparable to inert BHBs. Additionally, if there are no CCSN natal
kicks, inert NSBs are formed more efficiently on isolated fields than
in open clusters. Thus, we have concluded that open clusters cannot
form both Gaia BHs and NSs. However, we do not mean that open clusters
cannot form Gaia BHs. It would be possible that Gaia BHs are formed in
open clusters, while Gaia NSs are formed on isolated fields. We will
make it clear when more many Gaia BHs will be discovered in the
future.

\section*{Acknowledgments}

This research could not be accomplished without the support by
Grants-in-Aid for Scientific Research (19K03907, 24K07040) from the
Japan Society for the Promotion of Science.  M.F. is supported by The
University of Tokyo Excellent Young Researcher Program. L.W. thanks
the support from the one-hundred-talent project of Sun Yat-sen
University, the Fundamental Research Funds for the Central
Universities, Sun Yat-sen University (22hytd09) and the National
Natural Science Foundation of China through grant 12073090 and
12233013. Numerical simulations are carried out on Small Parallel
Computers at Center for Computational Astrophysics, National
Astronomical Observatory of Japan, the Yukawa Institute Computer
Facility, and Cygnus/Pegasus at the CCS, University of Tsukuba.

\section*{Data availability}

Results will be shared on reasonable request to authors. The data is
generated by the software {\tt PeTar} and {\tt SDAR}, which are
available in GitHub at \url{https://github.com/lwang-astro/PeTar} and
\url{https://github.com/lwang-astro/SDAR}, respectively. The initial
conditions of star cluster models are generated by the software {\tt
  MCLUSTER} \citep{2011MNRAS.417.2300K}, which is available in GitHub
at \url{https://github.com/lwang-astro/mcluster}.


\begin{thebibliography}{}
\makeatletter
\relax
\def\mn@urlcharsother{\let\do\@makeother \do\$\do\&\do\#\do\^\do\_\do\%\do\~}
\def\mn@doi{\begingroup\mn@urlcharsother \@ifnextchar [ {\mn@doi@}
  {\mn@doi@[]}}
\def\mn@doi@[#1]#2{\def\@tempa{#1}\ifx\@tempa\@empty \href
  {http://dx.doi.org/#2} {doi:#2}\else \href {http://dx.doi.org/#2} {#1}\fi
  \endgroup}
\def\mn@eprint#1#2{\mn@eprint@#1:#2::\@nil}
\def\mn@eprint@arXiv#1{\href {http://arxiv.org/abs/#1} {{\tt arXiv:#1}}}
\def\mn@eprint@dblp#1{\href {http://dblp.uni-trier.de/rec/bibtex/#1.xml}
  {dblp:#1}}
\def\mn@eprint@#1:#2:#3:#4\@nil{\def\@tempa {#1}\def\@tempb {#2}\def\@tempc
  {#3}\ifx \@tempc \@empty \let \@tempc \@tempb \let \@tempb \@tempa \fi \ifx
  \@tempb \@empty \def\@tempb {arXiv}\fi \@ifundefined
  {mn@eprint@\@tempb}{\@tempb:\@tempc}{\expandafter \expandafter \csname
  mn@eprint@\@tempb\endcsname \expandafter{\@tempc}}}

\bibitem[\protect\citeauthoryear{{Abbott} et~al.,}{{Abbott}
  et~al.}{2019}]{2019PhRvX...9c1040A}
{Abbott} B.~P.,  et~al., 2019, \mn@doi [Physical Review X]
  {10.1103/PhysRevX.9.031040}, \href
  {https://ui.adsabs.harvard.edu/abs/2019PhRvX...9c1040A} {9, 031040}

\bibitem[\protect\citeauthoryear{{Abbott} et~al.,}{{Abbott}
  et~al.}{2021}]{2021PhRvX..11b1053A}
{Abbott} R.,  et~al., 2021, \mn@doi [Physical Review X]
  {10.1103/PhysRevX.11.021053}, \href
  {https://ui.adsabs.harvard.edu/abs/2021PhRvX..11b1053A} {11, 021053}

\bibitem[\protect\citeauthoryear{{Abbott} et~al.,}{{Abbott}
  et~al.}{2023}]{2023PhRvX..13a1048A}
{Abbott} R.,  et~al., 2023, \mn@doi [Physical Review X]
  {10.1103/PhysRevX.13.011048}, \href
  {https://ui.adsabs.harvard.edu/abs/2023PhRvX..13a1048A} {13, 011048}

\bibitem[\protect\citeauthoryear{{Abdikamalov}, {Ott}, {Rezzolla}, {Dessart},
  {Dimmelmeier}, {Marek}  \& {Janka}}{{Abdikamalov}
  et~al.}{2010}]{2010PhRvD..81d4012A}
{Abdikamalov} E.~B.,  {Ott} C.~D.,  {Rezzolla} L.,  {Dessart} L.,
  {Dimmelmeier} H.,  {Marek} A.,   {Janka} H.~T.,  2010, \mn@doi [\prd]
  {10.1103/PhysRevD.81.044012}, \href
  {https://ui.adsabs.harvard.edu/abs/2010PhRvD..81d4012A} {81, 044012}

\bibitem[\protect\citeauthoryear{{Abdul-Masih} et~al.,}{{Abdul-Masih}
  et~al.}{2020}]{2020Natur.580E..11A}
{Abdul-Masih} M.,  et~al., 2020, \mn@doi [\nat] {10.1038/s41586-020-2216-x},
  \href {https://ui.adsabs.harvard.edu/abs/2020Natur.580E..11A} {580, E11}

\bibitem[\protect\citeauthoryear{{Andrews}, {Taggart}  \& {Foley}}{{Andrews}
  et~al.}{2022}]{2022arXiv220700680A}
{Andrews} J.~J.,  {Taggart} K.,   {Foley} R.,  2022, arXiv e-prints, \href
  {https://ui.adsabs.harvard.edu/abs/2022arXiv220700680A} {p. arXiv:2207.00680}

\bibitem[\protect\citeauthoryear{{Balbinot} et~al.,}{{Balbinot}
  et~al.}{2024}]{2024arXiv240411604B}
{Balbinot} E.,  et~al., 2024, \mn@doi [arXiv e-prints]
  {10.48550/arXiv.2404.11604}, \href
  {https://ui.adsabs.harvard.edu/abs/2024arXiv240411604B} {p. arXiv:2404.11604}

\bibitem[\protect\citeauthoryear{{Banerjee}, {Belczynski}, {Fryer}, {Berczik},
  {Hurley}, {Spurzem}  \& {Wang}}{{Banerjee}
  et~al.}{2020}]{2020A&A...639A..41B}
{Banerjee} S.,  {Belczynski} K.,  {Fryer} C.~L.,  {Berczik} P.,  {Hurley}
  J.~R.,  {Spurzem} R.,   {Wang} L.,  2020, \mn@doi [\aap]
  {10.1051/0004-6361/201935332}, \href
  {https://ui.adsabs.harvard.edu/abs/2020A&A...639A..41B} {639, A41}

\bibitem[\protect\citeauthoryear{{Baron}, {Cooperstein}, {Kahana}  \&
  {Nomoto}}{{Baron} et~al.}{1987}]{1987ApJ...320..304B}
{Baron} E.,  {Cooperstein} J.,  {Kahana} S.,   {Nomoto} K.,  1987, \mn@doi
  [\apj] {10.1086/165542}, \href
  {https://ui.adsabs.harvard.edu/abs/1987ApJ...320..304B} {320, 304}

\bibitem[\protect\citeauthoryear{{Belloni}, {Askar}, {Giersz}, {Kroupa}  \&
  {Rocha-Pinto}}{{Belloni} et~al.}{2017}]{2017MNRAS.471.2812B}
{Belloni} D.,  {Askar} A.,  {Giersz} M.,  {Kroupa} P.,   {Rocha-Pinto} H.~J.,
  2017, \mn@doi [\mnras] {10.1093/mnras/stx1763}, \href
  {https://ui.adsabs.harvard.edu/abs/2017MNRAS.471.2812B} {471, 2812}

\bibitem[\protect\citeauthoryear{{Bodensteiner} et~al.,}{{Bodensteiner}
  et~al.}{2020}]{2020A&A...641A..43B}
{Bodensteiner} J.,  et~al., 2020, \mn@doi [\aap] {10.1051/0004-6361/202038682},
  \href {https://ui.adsabs.harvard.edu/abs/2020A&A...641A..43B} {641, A43}

\bibitem[\protect\citeauthoryear{{Bovy}}{{Bovy}}{2015}]{2015ApJS..216...29B}
{Bovy} J.,  2015, \mn@doi [\apjs] {10.1088/0067-0049/216/2/29}, \href
  {https://ui.adsabs.harvard.edu/abs/2015ApJS..216...29B} {216, 29}

\bibitem[\protect\citeauthoryear{{Casares}, {Jonker}  \& {Israelian}}{{Casares}
  et~al.}{2017}]{2017hsn..book.1499C}
{Casares} J.,  {Jonker} P.~G.,   {Israelian} G.,  2017, {X-Ray Binaries}.
p.~1499, \mn@doi{10.1007/978-3-319-21846-5_111}

\bibitem[\protect\citeauthoryear{{Chakrabarti} et~al.,}{{Chakrabarti}
  et~al.}{2023}]{2023AJ....166....6C}
{Chakrabarti} S.,  et~al., 2023, \mn@doi [\aj] {10.3847/1538-3881/accf21},
  \href {https://ui.adsabs.harvard.edu/abs/2023AJ....166....6C} {166, 6}

\bibitem[\protect\citeauthoryear{{Chawla}, {Chatterjee}, {Breivik}, {Moorthy},
  {Andrews}  \& {Sanderson}}{{Chawla} et~al.}{2022}]{2022ApJ...931..107C}
{Chawla} C.,  {Chatterjee} S.,  {Breivik} K.,  {Moorthy} C.~K.,  {Andrews}
  J.~J.,   {Sanderson} R.~E.,  2022, \mn@doi [\apj] {10.3847/1538-4357/ac60a5},
  \href {https://ui.adsabs.harvard.edu/abs/2022ApJ...931..107C} {931, 107}

\bibitem[\protect\citeauthoryear{{Claeys}, {Pols}, {Izzard}, {Vink}  \&
  {Verbunt}}{{Claeys} et~al.}{2014}]{2014A&A...563A..83C}
{Claeys} J.~S.~W.,  {Pols} O.~R.,  {Izzard} R.~G.,  {Vink} J.,   {Verbunt}
  F.~W.~M.,  2014, \mn@doi [\aap] {10.1051/0004-6361/201322714}, \href
  {https://ui.adsabs.harvard.edu/abs/2014A&A...563A..83C} {563, A83}

\bibitem[\protect\citeauthoryear{{Dan}, {Rosswog}, {Br{\"u}ggen}  \&
  {Podsiadlowski}}{{Dan} et~al.}{2014}]{2014MNRAS.438...14D}
{Dan} M.,  {Rosswog} S.,  {Br{\"u}ggen} M.,   {Podsiadlowski} P.,  2014,
  \mn@doi [\mnras] {10.1093/mnras/stt1766}, \href
  {https://ui.adsabs.harvard.edu/abs/2014MNRAS.438...14D} {438, 14}

\bibitem[\protect\citeauthoryear{{Dessart}, {Burrows}, {Ott}, {Livne}, {Yoon}
  \& {Langer}}{{Dessart} et~al.}{2006}]{2006ApJ...644.1063D}
{Dessart} L.,  {Burrows} A.,  {Ott} C.~D.,  {Livne} E.,  {Yoon} S.~C.,
  {Langer} N.,  2006, \mn@doi [\apj] {10.1086/503626}, \href
  {https://ui.adsabs.harvard.edu/abs/2006ApJ...644.1063D} {644, 1063}

\bibitem[\protect\citeauthoryear{{Di Carlo}, {Agrawal}, {Rodriguez}  \&
  {Breivik}}{{Di Carlo} et~al.}{2024}]{2024ApJ...965...22D}
{Di Carlo} U.~N.,  {Agrawal} P.,  {Rodriguez} C.~L.,   {Breivik} K.,  2024,
  \mn@doi [\apj] {10.3847/1538-4357/ad2f2c}, \href
  {https://ui.adsabs.harvard.edu/abs/2024ApJ...965...22D} {965, 22}

\bibitem[\protect\citeauthoryear{{El-Badry}}{{El-Badry}}{2024a}]{2024arXiv240312146E}
{El-Badry} K.,  2024a, \mn@doi [arXiv e-prints] {10.48550/arXiv.2403.12146},
  \href {https://ui.adsabs.harvard.edu/abs/2024arXiv240312146E} {p.
  arXiv:2403.12146}

\bibitem[\protect\citeauthoryear{{El-Badry}}{{El-Badry}}{2024b}]{2024arXiv240413047E}
{El-Badry} K.,  2024b, \mn@doi [arXiv e-prints] {10.48550/arXiv.2404.13047},
  \href {https://ui.adsabs.harvard.edu/abs/2024arXiv240413047E} {p.
  arXiv:2404.13047}

\bibitem[\protect\citeauthoryear{{El-Badry} \& {Burdge}}{{El-Badry} \&
  {Burdge}}{2022}]{2022MNRAS.511L..24E}
{El-Badry} K.,  {Burdge} K.~B.,  2022, \mn@doi [\mnras]
  {10.1093/mnrasl/slab135}, \href
  {https://ui.adsabs.harvard.edu/abs/2022MNRAS.511L..24E} {511, 24}

\bibitem[\protect\citeauthoryear{{El-Badry} \& {Quataert}}{{El-Badry} \&
  {Quataert}}{2020}]{2020MNRAS.493L..22E}
{El-Badry} K.,  {Quataert} E.,  2020, \mn@doi [\mnras]
  {10.1093/mnrasl/slaa004}, \href
  {https://ui.adsabs.harvard.edu/abs/2020MNRAS.493L..22E} {493, L22}

\bibitem[\protect\citeauthoryear{{El-Badry} \& {Quataert}}{{El-Badry} \&
  {Quataert}}{2021}]{2021MNRAS.502.3436E}
{El-Badry} K.,  {Quataert} E.,  2021, \mn@doi [\mnras] {10.1093/mnras/stab285},
  \href {https://ui.adsabs.harvard.edu/abs/2021MNRAS.502.3436E} {502, 3436}

\bibitem[\protect\citeauthoryear{{El-Badry}, {Burdge}  \&
  {Mr{\'o}z}}{{El-Badry} et~al.}{2022a}]{2022MNRAS.511.3089E}
{El-Badry} K.,  {Burdge} K.~B.,   {Mr{\'o}z} P.,  2022a, \mn@doi [\mnras]
  {10.1093/mnras/stac274}, \href
  {https://ui.adsabs.harvard.edu/abs/2022MNRAS.511.3089E} {511, 3089}

\bibitem[\protect\citeauthoryear{{El-Badry}, {Seeburger}, {Jayasinghe}, {Rix},
  {Almada}, {Conroy}, {Price-Whelan}  \& {Burdge}}{{El-Badry}
  et~al.}{2022b}]{2022MNRAS.512.5620E}
{El-Badry} K.,  {Seeburger} R.,  {Jayasinghe} T.,  {Rix} H.-W.,  {Almada} S.,
  {Conroy} C.,  {Price-Whelan} A.~M.,   {Burdge} K.,  2022b, \mn@doi [\mnras]
  {10.1093/mnras/stac815}, \href
  {https://ui.adsabs.harvard.edu/abs/2022MNRAS.512.5620E} {512, 5620}

\bibitem[\protect\citeauthoryear{{El-Badry} et~al.,}{{El-Badry}
  et~al.}{2023a}]{2023OJAp....6E..28E}
{El-Badry} K.,  et~al., 2023a, \mn@doi [The Open Journal of Astrophysics]
  {10.21105/astro.2306.03914}, \href
  {https://ui.adsabs.harvard.edu/abs/2023OJAp....6E..28E} {6, 28}

\bibitem[\protect\citeauthoryear{{El-Badry} et~al.,}{{El-Badry}
  et~al.}{2023b}]{2023MNRAS.518.1057E}
{El-Badry} K.,  et~al., 2023b, \mn@doi [\mnras] {10.1093/mnras/stac3140}, \href
  {https://ui.adsabs.harvard.edu/abs/2023MNRAS.518.1057E} {518, 1057}

\bibitem[\protect\citeauthoryear{{El-Badry} et~al.,}{{El-Badry}
  et~al.}{2023c}]{2023MNRAS.521.4323E}
{El-Badry} K.,  et~al., 2023c, \mn@doi [\mnras] {10.1093/mnras/stad799}, \href
  {https://ui.adsabs.harvard.edu/abs/2023MNRAS.521.4323E} {521, 4323}

\bibitem[\protect\citeauthoryear{{El-Badry} et~al.,}{{El-Badry}
  et~al.}{2024a}]{2024arXiv240500089E}
{El-Badry} K.,  et~al., 2024a, \mn@doi [arXiv e-prints]
  {10.48550/arXiv.2405.00089}, \href
  {https://ui.adsabs.harvard.edu/abs/2024arXiv240500089E} {p. arXiv:2405.00089}

\bibitem[\protect\citeauthoryear{{El-Badry} et~al.,}{{El-Badry}
  et~al.}{2024b}]{2024OJAp....7E..27E}
{El-Badry} K.,  et~al., 2024b, \mn@doi [The Open Journal of Astrophysics]
  {10.33232/001c.116675}, \href
  {https://ui.adsabs.harvard.edu/abs/2024OJAp....7E..27E} {7, 27}

\bibitem[\protect\citeauthoryear{{Ferrand}, {Tanikawa}, {Warren}, {Nagataki},
  {Safi-Harb}  \& {Decourchelle}}{{Ferrand} et~al.}{2022}]{2022ApJ...930...92F}
{Ferrand} G.,  {Tanikawa} A.,  {Warren} D.~C.,  {Nagataki} S.,  {Safi-Harb} S.,
    {Decourchelle} A.,  2022, \mn@doi [\apj] {10.3847/1538-4357/ac5c58}, \href
  {https://ui.adsabs.harvard.edu/abs/2022ApJ...930...92F} {930, 92}

\bibitem[\protect\citeauthoryear{{Fryer}, {Benz}, {Herant}  \&
  {Colgate}}{{Fryer} et~al.}{1999}]{1999ApJ...516..892F}
{Fryer} C.,  {Benz} W.,  {Herant} M.,   {Colgate} S.~A.,  1999, \mn@doi [\apj]
  {10.1086/307119}, \href
  {https://ui.adsabs.harvard.edu/abs/1999ApJ...516..892F} {516, 892}

\bibitem[\protect\citeauthoryear{{Fryer}, {Belczynski}, {Wiktorowicz},
  {Dominik}, {Kalogera}  \& {Holz}}{{Fryer} et~al.}{2012}]{2012ApJ...749...91F}
{Fryer} C.~L.,  {Belczynski} K.,  {Wiktorowicz} G.,  {Dominik} M.,  {Kalogera}
  V.,   {Holz} D.~E.,  2012, \mn@doi [\apj] {10.1088/0004-637X/749/1/91}, \href
  {https://ui.adsabs.harvard.edu/abs/2012ApJ...749...91F} {749, 91}

\bibitem[\protect\citeauthoryear{{Gaia Collaboration} et~al.,}{{Gaia
  Collaboration} et~al.}{2023a}]{2023A&A...674A...1G}
{Gaia Collaboration} et~al., 2023a, \mn@doi [\aap]
  {10.1051/0004-6361/202243940}, \href
  {https://ui.adsabs.harvard.edu/abs/2023A&A...674A...1G} {674, A1}

\bibitem[\protect\citeauthoryear{{Gaia Collaboration} et~al.,}{{Gaia
  Collaboration} et~al.}{2023b}]{2023A&A...674A..34G}
{Gaia Collaboration} et~al., 2023b, \mn@doi [\aap]
  {10.1051/0004-6361/202243782}, \href
  {https://ui.adsabs.harvard.edu/abs/2023A&A...674A..34G} {674, A34}

\bibitem[\protect\citeauthoryear{{Gaia Collaboration} et~al.,}{{Gaia
  Collaboration} et~al.}{2024}]{2024arXiv240410486G}
{Gaia Collaboration} et~al., 2024, \mn@doi [arXiv e-prints]
  {10.48550/arXiv.2404.10486}, \href
  {https://ui.adsabs.harvard.edu/abs/2024arXiv240410486G} {p. arXiv:2404.10486}

\bibitem[\protect\citeauthoryear{{Geier}, {Dorsch}, {Dawson}, {Pelisoli},
  {Munday}, {Marsh}, {Schaffenroth}  \& {Heber}}{{Geier}
  et~al.}{2023}]{2023A&A...677A..11G}
{Geier} S.,  {Dorsch} M.,  {Dawson} H.,  {Pelisoli} I.,  {Munday} J.,  {Marsh}
  T.~R.,  {Schaffenroth} V.,   {Heber} U.,  2023, \mn@doi [\aap]
  {10.1051/0004-6361/202346407}, \href
  {https://ui.adsabs.harvard.edu/abs/2023A&A...677A..11G} {677, A11}

\bibitem[\protect\citeauthoryear{{Generozov} \& {Perets}}{{Generozov} \&
  {Perets}}{2023}]{2023arXiv231203066G}
{Generozov} A.,  {Perets} H.~B.,  2023, \mn@doi [arXiv e-prints]
  {10.48550/arXiv.2312.03066}, \href
  {https://ui.adsabs.harvard.edu/abs/2023arXiv231203066G} {p. arXiv:2312.03066}

\bibitem[\protect\citeauthoryear{{Gessner} \& {Janka}}{{Gessner} \&
  {Janka}}{2018}]{2018ApJ...865...61G}
{Gessner} A.,  {Janka} H.-T.,  2018, \mn@doi [\apj] {10.3847/1538-4357/aadbae},
  \href {https://ui.adsabs.harvard.edu/abs/2018ApJ...865...61G} {865, 61}

\bibitem[\protect\citeauthoryear{{Giesers} et~al.,}{{Giesers}
  et~al.}{2018}]{2018MNRAS.475L..15G}
{Giesers} B.,  et~al., 2018, \mn@doi [\mnras] {10.1093/mnrasl/slx203}, \href
  {https://ui.adsabs.harvard.edu/abs/2018MNRAS.475L..15G} {475, L15}

\bibitem[\protect\citeauthoryear{{Gronow}, {Collins}, {Ohlmann}, {Pakmor},
  {Kromer}, {Seitenzahl}, {Sim}  \& {R{\"o}pke}}{{Gronow}
  et~al.}{2020}]{2020A&A...635A.169G}
{Gronow} S.,  {Collins} C.,  {Ohlmann} S.~T.,  {Pakmor} R.,  {Kromer} M.,
  {Seitenzahl} I.~R.,  {Sim} S.~A.,   {R{\"o}pke} F.~K.,  2020, \mn@doi [\aap]
  {10.1051/0004-6361/201936494}, \href
  {https://ui.adsabs.harvard.edu/abs/2020A&A...635A.169G} {635, A169}

\bibitem[\protect\citeauthoryear{{Guillochon}, {Dan}, {Ramirez-Ruiz}  \&
  {Rosswog}}{{Guillochon} et~al.}{2010}]{2010ApJ...709L..64G}
{Guillochon} J.,  {Dan} M.,  {Ramirez-Ruiz} E.,   {Rosswog} S.,  2010, \mn@doi
  [\apjl] {10.1088/2041-8205/709/1/L64}, \href
  {http://ads.nao.ac.jp/abs/2010ApJ...709L..64G} {709, L64}

\bibitem[\protect\citeauthoryear{{Heggie}}{{Heggie}}{1975}]{1975MNRAS.173..729H}
{Heggie} D.~C.,  1975, \mn@doi [\mnras] {10.1093/mnras/173.3.729}, \href
  {http://adsabs.harvard.edu/abs/1975MNRAS.173..729H} {173, 729}

\bibitem[\protect\citeauthoryear{{Hirai} \& {Mandel}}{{Hirai} \&
  {Mandel}}{2022}]{2022ApJ...937L..42H}
{Hirai} R.,  {Mandel} I.,  2022, \mn@doi [\apjl] {10.3847/2041-8213/ac9519},
  \href {https://ui.adsabs.harvard.edu/abs/2022ApJ...937L..42H} {937, L42}

\bibitem[\protect\citeauthoryear{{Hobbs}, {Lorimer}, {Lyne}  \&
  {Kramer}}{{Hobbs} et~al.}{2005}]{2005MNRAS.360..974H}
{Hobbs} G.,  {Lorimer} D.~R.,  {Lyne} A.~G.,   {Kramer} M.,  2005, \mn@doi
  [\mnras] {10.1111/j.1365-2966.2005.09087.x}, \href
  {https://ui.adsabs.harvard.edu/abs/2005MNRAS.360..974H} {360, 974}

\bibitem[\protect\citeauthoryear{Howil et~al.,}{Howil
  et~al.}{2024}]{howil2024uncovering}
Howil K.,  et~al., 2024, Uncovering the Invisible: A Study of Gaia18ajz, a
  Candidate Black Hole Revealed by Microlensing (\mn@eprint {arXiv}
  {2403.09006})

\bibitem[\protect\citeauthoryear{{Hurley}, {Pols}  \& {Tout}}{{Hurley}
  et~al.}{2000}]{2000MNRAS.315..543H}
{Hurley} J.~R.,  {Pols} O.~R.,   {Tout} C.~A.,  2000, \mn@doi [\mnras]
  {10.1046/j.1365-8711.2000.03426.x}, \href
  {http://adsabs.harvard.edu/abs/2000MNRAS.315..543H} {315, 543}

\bibitem[\protect\citeauthoryear{{Hurley}, {Tout}  \& {Pols}}{{Hurley}
  et~al.}{2002}]{2002MNRAS.329..897H}
{Hurley} J.~R.,  {Tout} C.~A.,   {Pols} O.~R.,  2002, \mn@doi [\mnras]
  {10.1046/j.1365-8711.2002.05038.x}, \href
  {http://adsabs.harvard.edu/abs/2002MNRAS.329..897H} {329, 897}

\bibitem[\protect\citeauthoryear{{Iorio} et~al.,}{{Iorio}
  et~al.}{2024}]{2024arXiv240417568I}
{Iorio} G.,  et~al., 2024, \mn@doi [arXiv e-prints]
  {10.48550/arXiv.2404.17568}, \href
  {https://ui.adsabs.harvard.edu/abs/2024arXiv240417568I} {p. arXiv:2404.17568}

\bibitem[\protect\citeauthoryear{{Ivanova}, {Heinke}, {Rasio}, {Belczynski}  \&
  {Fregeau}}{{Ivanova} et~al.}{2008}]{2008MNRAS.386..553I}
{Ivanova} N.,  {Heinke} C.~O.,  {Rasio} F.~A.,  {Belczynski} K.,   {Fregeau}
  J.~M.,  2008, \mn@doi [\mnras] {10.1111/j.1365-2966.2008.13064.x}, \href
  {https://ui.adsabs.harvard.edu/abs/2008MNRAS.386..553I} {386, 553}

\bibitem[\protect\citeauthoryear{{Iwasawa}, {Tanikawa}, {Hosono}, {Nitadori},
  {Muranushi}  \& {Makino}}{{Iwasawa} et~al.}{2016}]{2016PASJ...68...54I}
{Iwasawa} M.,  {Tanikawa} A.,  {Hosono} N.,  {Nitadori} K.,  {Muranushi} T.,
  {Makino} J.,  2016, \mn@doi [\pasj] {10.1093/pasj/psw053}, \href
  {http://ads.nao.ac.jp/abs/2016PASJ...68...54I} {68, 54}

\bibitem[\protect\citeauthoryear{{Iwasawa}, {Oshino}, {Fujii}  \&
  {Hori}}{{Iwasawa} et~al.}{2017}]{2017PASJ...69...81I}
{Iwasawa} M.,  {Oshino} S.,  {Fujii} M.~S.,   {Hori} Y.,  2017, \mn@doi [\pasj]
  {10.1093/pasj/psx073}, \href
  {https://ui.adsabs.harvard.edu/abs/2017PASJ...69...81I} {69, 81}

\bibitem[\protect\citeauthoryear{{Iwasawa}, {Namekata}, {Nitadori}, {Nomura},
  {Wang}, {Tsubouchi}  \& {Makino}}{{Iwasawa}
  et~al.}{2020}]{2020PASJ...72...13I}
{Iwasawa} M.,  {Namekata} D.,  {Nitadori} K.,  {Nomura} K.,  {Wang} L.,
  {Tsubouchi} M.,   {Makino} J.,  2020, \mn@doi [\pasj] {10.1093/pasj/psz133},
  \href {https://ui.adsabs.harvard.edu/abs/2020PASJ...72...13I} {72, 13}

\bibitem[\protect\citeauthoryear{{Jayasinghe} et~al.,}{{Jayasinghe}
  et~al.}{2021}]{2021MNRAS.504.2577J}
{Jayasinghe} T.,  et~al., 2021, \mn@doi [\mnras] {10.1093/mnras/stab907}, \href
  {https://ui.adsabs.harvard.edu/abs/2021MNRAS.504.2577J} {504, 2577}

\bibitem[\protect\citeauthoryear{{Jayasinghe} et~al.,}{{Jayasinghe}
  et~al.}{2022}]{2022MNRAS.516.5945J}
{Jayasinghe} T.,  et~al., 2022, \mn@doi [\mnras] {10.1093/mnras/stac2187},
  \href {https://ui.adsabs.harvard.edu/abs/2022MNRAS.516.5945J} {516, 5945}

\bibitem[\protect\citeauthoryear{{Jayasinghe}, {Rowan}, {Thompson}, {Kochanek}
  \& {Stanek}}{{Jayasinghe} et~al.}{2023}]{2023MNRAS.521.5927J}
{Jayasinghe} T.,  {Rowan} D.~M.,  {Thompson} T.~A.,  {Kochanek} C.~S.,
  {Stanek} K.~Z.,  2023, \mn@doi [\mnras] {10.1093/mnras/stad909}, \href
  {https://ui.adsabs.harvard.edu/abs/2023MNRAS.521.5927J} {521, 5927}

\bibitem[\protect\citeauthoryear{{Jones} et~al.,}{{Jones}
  et~al.}{2013}]{2013ApJ...772..150J}
{Jones} S.,  et~al., 2013, \mn@doi [\apj] {10.1088/0004-637X/772/2/150}, \href
  {https://ui.adsabs.harvard.edu/abs/2013ApJ...772..150J} {772, 150}

\bibitem[\protect\citeauthoryear{{Kashyap}, {Fisher}, {Garc{\'{\i}}a-Berro},
  {Aznar-Sigu{\'a}n}, {Ji}  \& {Lor{\'e}n-Aguilar}}{{Kashyap}
  et~al.}{2015}]{2015ApJ...800L...7K}
{Kashyap} R.,  {Fisher} R.,  {Garc{\'{\i}}a-Berro} E.,  {Aznar-Sigu{\'a}n} G.,
  {Ji} S.,   {Lor{\'e}n-Aguilar} P.,  2015, \mn@doi [\apjl]
  {10.1088/2041-8205/800/1/L7}, \href
  {http://adsabs.harvard.edu/abs/2015ApJ...800L...7K} {800, L7}

\bibitem[\protect\citeauthoryear{{Kashyap}, {Haque}, {Lor{\'e}n-Aguilar},
  {Garc{\'\i}a-Berro}  \& {Fisher}}{{Kashyap}
  et~al.}{2018}]{2018ApJ...869..140K}
{Kashyap} R.,  {Haque} T.,  {Lor{\'e}n-Aguilar} P.,  {Garc{\'\i}a-Berro} E.,
  {Fisher} R.,  2018, \mn@doi [\apj] {10.3847/1538-4357/aaedb7}, \href
  {https://ui.adsabs.harvard.edu/abs/2018ApJ...869..140K} {869, 140}

\bibitem[\protect\citeauthoryear{{Kotko}, {Banerjee}  \& {Belczynski}}{{Kotko}
  et~al.}{2024}]{2024arXiv240313579K}
{Kotko} I.,  {Banerjee} S.,   {Belczynski} K.,  2024, arXiv e-prints, \href
  {https://ui.adsabs.harvard.edu/abs/2024arXiv240313579K} {p. arXiv:2403.13579}

\bibitem[\protect\citeauthoryear{{Kremer}, {Ye}, {Chatterjee}, {Rodriguez}  \&
  {Rasio}}{{Kremer} et~al.}{2018}]{2018ApJ...855L..15K}
{Kremer} K.,  {Ye} C.~S.,  {Chatterjee} S.,  {Rodriguez} C.~L.,   {Rasio}
  F.~A.,  2018, \mn@doi [\apjl] {10.3847/2041-8213/aab26c}, \href
  {https://ui.adsabs.harvard.edu/abs/2018ApJ...855L..15K} {855, L15}

\bibitem[\protect\citeauthoryear{{Kremer} et~al.,}{{Kremer}
  et~al.}{2020}]{2020ApJS..247...48K}
{Kremer} K.,  et~al., 2020, \mn@doi [\apjs] {10.3847/1538-4365/ab7919}, \href
  {https://ui.adsabs.harvard.edu/abs/2020ApJS..247...48K} {247, 48}

\bibitem[\protect\citeauthoryear{{Kremer}, {Fuller}, {Piro}  \&
  {Ransom}}{{Kremer} et~al.}{2023}]{2023MNRAS.525L..22K}
{Kremer} K.,  {Fuller} J.,  {Piro} A.~L.,   {Ransom} S.~M.,  2023, \mn@doi
  [\mnras] {10.1093/mnrasl/slad088}, \href
  {https://ui.adsabs.harvard.edu/abs/2023MNRAS.525L..22K} {525, L22}

\bibitem[\protect\citeauthoryear{{Kroupa}}{{Kroupa}}{1995a}]{1995MNRAS.277.1491K}
{Kroupa} P.,  1995a, \mn@doi [\mnras] {10.1093/mnras/277.4.1491}, \href
  {https://ui.adsabs.harvard.edu/abs/1995MNRAS.277.1491K} {277, 1491}

\bibitem[\protect\citeauthoryear{{Kroupa}}{{Kroupa}}{1995b}]{1995MNRAS.277.1507K}
{Kroupa} P.,  1995b, \mn@doi [\mnras] {10.1093/mnras/277.4.1507}, \href
  {https://ui.adsabs.harvard.edu/abs/1995MNRAS.277.1507K} {277, 1507}

\bibitem[\protect\citeauthoryear{{Kroupa}}{{Kroupa}}{2001}]{2001MNRAS.322..231K}
{Kroupa} P.,  2001, \mn@doi [\mnras] {10.1046/j.1365-8711.2001.04022.x}, \href
  {http://adsabs.harvard.edu/abs/2001MNRAS.322..231K} {322, 231}

\bibitem[\protect\citeauthoryear{{K{\"u}pper}, {Maschberger}, {Kroupa}  \&
  {Baumgardt}}{{K{\"u}pper} et~al.}{2011}]{2011MNRAS.417.2300K}
{K{\"u}pper} A. H.~W.,  {Maschberger} T.,  {Kroupa} P.,   {Baumgardt} H.,
  2011, \mn@doi [\mnras] {10.1111/j.1365-2966.2011.19412.x}, \href
  {https://ui.adsabs.harvard.edu/abs/2011MNRAS.417.2300K} {417, 2300}

\bibitem[\protect\citeauthoryear{{Lam} et~al.,}{{Lam}
  et~al.}{2022}]{2022ApJ...933L..23L}
{Lam} C.~Y.,  et~al., 2022, \mn@doi [\apjl] {10.3847/2041-8213/ac7442}, \href
  {https://ui.adsabs.harvard.edu/abs/2022ApJ...933L..23L} {933, L23}

\bibitem[\protect\citeauthoryear{{Lennon}, {Dufton}, {Villase{\~n}or}, {Evans},
  {Langer}, {Saxton}, {Monageng}  \& {Toonen}}{{Lennon}
  et~al.}{2022}]{2022A&A...665A.180L}
{Lennon} D.~J.,  {Dufton} P.~L.,  {Villase{\~n}or} J.~I.,  {Evans} C.~J.,
  {Langer} N.,  {Saxton} R.,  {Monageng} I.~M.,   {Toonen} S.,  2022, \mn@doi
  [\aap] {10.1051/0004-6361/202142413}, \href
  {https://ui.adsabs.harvard.edu/abs/2022A&A...665A.180L} {665, A180}

\bibitem[\protect\citeauthoryear{{Liu} et~al.,}{{Liu}
  et~al.}{2019}]{2019Natur.575..618L}
{Liu} J.,  et~al., 2019, \mn@doi [\nat] {10.1038/s41586-019-1766-2}, \href
  {https://ui.adsabs.harvard.edu/abs/2019Natur.575..618L} {575, 618}

\bibitem[\protect\citeauthoryear{{Longo Micchi}, {Radice}  \&
  {Chirenti}}{{Longo Micchi} et~al.}{2023}]{2023MNRAS.525.6359L}
{Longo Micchi} L.~F.,  {Radice} D.,   {Chirenti} C.,  2023, \mn@doi [\mnras]
  {10.1093/mnras/stad2420}, \href
  {https://ui.adsabs.harvard.edu/abs/2023MNRAS.525.6359L} {525, 6359}

\bibitem[\protect\citeauthoryear{{Mar{\'\i}n Pina}, {Rastello}, {Gieles},
  {Kremer}, {Fitzgerald}  \& {Rando}}{{Mar{\'\i}n Pina}
  et~al.}{2024}]{2024arXiv240413036M}
{Mar{\'\i}n Pina} D.,  {Rastello} S.,  {Gieles} M.,  {Kremer} K.,  {Fitzgerald}
  L.,   {Rando} B.,  2024, \mn@doi [arXiv e-prints]
  {10.48550/arXiv.2404.13036}, \href
  {https://ui.adsabs.harvard.edu/abs/2024arXiv240413036M} {p. arXiv:2404.13036}

\bibitem[\protect\citeauthoryear{{Miyaji}, {Nomoto}, {Yokoi}  \&
  {Sugimoto}}{{Miyaji} et~al.}{1980}]{1980PASJ...32..303M}
{Miyaji} S.,  {Nomoto} K.,  {Yokoi} K.,   {Sugimoto} D.,  1980, \pasj, \href
  {https://ui.adsabs.harvard.edu/abs/1980PASJ...32..303M} {32, 303}

\bibitem[\protect\citeauthoryear{{Moe} \& {Di Stefano}}{{Moe} \& {Di
  Stefano}}{2017}]{2017ApJS..230...15M}
{Moe} M.,  {Di Stefano} R.,  2017, \mn@doi [\apjs] {10.3847/1538-4365/aa6fb6},
  \href {https://ui.adsabs.harvard.edu/abs/2017ApJS..230...15M} {230, 15}

\bibitem[\protect\citeauthoryear{{Mori}, {Sawada}, {Suwa}, {Tanikawa},
  {Kashiyama}  \& {Murase}}{{Mori} et~al.}{2023}]{2023arXiv230617381M}
{Mori} M.,  {Sawada} R.,  {Suwa} Y.,  {Tanikawa} A.,  {Kashiyama} K.,
  {Murase} K.,  2023, \mn@doi [arXiv e-prints] {10.48550/arXiv.2306.17381},
  \href {https://ui.adsabs.harvard.edu/abs/2023arXiv230617381M} {p.
  arXiv:2306.17381}

\bibitem[\protect\citeauthoryear{{Nomoto}}{{Nomoto}}{1982}]{1982ApJ...257..780N}
{Nomoto} K.,  1982, \mn@doi [\apj] {10.1086/160031}, \href
  {http://ads.nao.ac.jp/abs/1982ApJ...257..780N} {257, 780}

\bibitem[\protect\citeauthoryear{{Nomoto}}{{Nomoto}}{1984}]{1984ApJ...277..791N}
{Nomoto} K.,  1984, \mn@doi [\apj] {10.1086/161749}, \href
  {https://ui.adsabs.harvard.edu/abs/1984ApJ...277..791N} {277, 791}

\bibitem[\protect\citeauthoryear{{Nomoto}}{{Nomoto}}{1987}]{1987ApJ...322..206N}
{Nomoto} K.,  1987, \mn@doi [\apj] {10.1086/165716}, \href
  {https://ui.adsabs.harvard.edu/abs/1987ApJ...322..206N} {322, 206}

\bibitem[\protect\citeauthoryear{{Nomoto} \& {Kondo}}{{Nomoto} \&
  {Kondo}}{1991}]{1991ApJ...367L..19N}
{Nomoto} K.,  {Kondo} Y.,  1991, \mn@doi [\apjl] {10.1086/185922}, \href
  {https://ui.adsabs.harvard.edu/abs/1991ApJ...367L..19N} {367, L19}

\bibitem[\protect\citeauthoryear{{Oshino}, {Funato}  \& {Makino}}{{Oshino}
  et~al.}{2011}]{2011PASJ...63..881O}
{Oshino} S.,  {Funato} Y.,   {Makino} J.,  2011, \mn@doi [\pasj]
  {10.1093/pasj/63.4.881}, \href
  {https://ui.adsabs.harvard.edu/abs/2011PASJ...63..881O} {63, 881}

\bibitem[\protect\citeauthoryear{{Pakmor}, {Kromer}, {R{\"o}pke}, {Sim},
  {Ruiter}  \& {Hillebrandt}}{{Pakmor} et~al.}{2010}]{2010Natur.463...61P}
{Pakmor} R.,  {Kromer} M.,  {R{\"o}pke} F.~K.,  {Sim} S.~A.,  {Ruiter} A.~J.,
  {Hillebrandt} W.,  2010, \mn@doi [\nat] {10.1038/nature08642}, \href
  {http://ads.nao.ac.jp/abs/2010Natur.463...61P} {463, 61}

\bibitem[\protect\citeauthoryear{{Pakmor}, {Hachinger}, {R{\"o}pke}  \&
  {Hillebrandt}}{{Pakmor} et~al.}{2011}]{2011A&A...528A.117P}
{Pakmor} R.,  {Hachinger} S.,  {R{\"o}pke} F.~K.,   {Hillebrandt} W.,  2011,
  \mn@doi [\aap] {10.1051/0004-6361/201015653}, \href
  {https://ui.adsabs.harvard.edu/abs/2011A&A...528A.117P} {528, A117}

\bibitem[\protect\citeauthoryear{{Pakmor}, {Edelmann}, {R{\"o}pke}  \&
  {Hillebrandt}}{{Pakmor} et~al.}{2012a}]{2012MNRAS.424.2222P}
{Pakmor} R.,  {Edelmann} P.,  {R{\"o}pke} F.~K.,   {Hillebrandt} W.,  2012a,
  \mn@doi [\mnras] {10.1111/j.1365-2966.2012.21383.x}, \href
  {http://ads.nao.ac.jp/abs/2012MNRAS.424.2222P} {424, 2222}

\bibitem[\protect\citeauthoryear{{Pakmor}, {Kromer}, {Taubenberger}, {Sim},
  {R{\"o}pke}  \& {Hillebrandt}}{{Pakmor} et~al.}{2012b}]{2012ApJ...747L..10P}
{Pakmor} R.,  {Kromer} M.,  {Taubenberger} S.,  {Sim} S.~A.,  {R{\"o}pke}
  F.~K.,   {Hillebrandt} W.,  2012b, \mn@doi [\apjl]
  {10.1088/2041-8205/747/1/L10}, \href
  {https://ui.adsabs.harvard.edu/abs/2012ApJ...747L..10P} {747, L10}

\bibitem[\protect\citeauthoryear{{Pakmor}, {Kromer}, {Taubenberger}  \&
  {Springel}}{{Pakmor} et~al.}{2013}]{2013ApJ...770L...8P}
{Pakmor} R.,  {Kromer} M.,  {Taubenberger} S.,   {Springel} V.,  2013, \mn@doi
  [\apjl] {10.1088/2041-8205/770/1/L8}, \href
  {http://ads.nao.ac.jp/abs/2013ApJ...770L...8P} {770, L8}

\bibitem[\protect\citeauthoryear{{Pakmor}, {Zenati}, {Perets}  \&
  {Toonen}}{{Pakmor} et~al.}{2021}]{2021MNRAS.503.4734P}
{Pakmor} R.,  {Zenati} Y.,  {Perets} H.~B.,   {Toonen} S.,  2021, \mn@doi
  [\mnras] {10.1093/mnras/stab686}, \href
  {https://ui.adsabs.harvard.edu/abs/2021MNRAS.503.4734P} {503, 4734}

\bibitem[\protect\citeauthoryear{{Perets} et~al.,}{{Perets}
  et~al.}{2010}]{2010Natur.465..322P}
{Perets} H.~B.,  et~al., 2010, \mn@doi [\nat] {10.1038/nature09056}, \href
  {http://ads.nao.ac.jp/abs/2010Natur.465..322P} {465, 322}

\bibitem[\protect\citeauthoryear{{Philippov} \& {Kramer}}{{Philippov} \&
  {Kramer}}{2022}]{2022ARA&A..60..495P}
{Philippov} A.,  {Kramer} M.,  2022, \mn@doi [\araa]
  {10.1146/annurev-astro-052920-112338}, \href
  {https://ui.adsabs.harvard.edu/abs/2022ARA&A..60..495P} {60, 495}

\bibitem[\protect\citeauthoryear{{Podsiadlowski}, {Langer}, {Poelarends},
  {Rappaport}, {Heger}  \& {Pfahl}}{{Podsiadlowski}
  et~al.}{2004}]{2004ApJ...612.1044P}
{Podsiadlowski} P.,  {Langer} N.,  {Poelarends} A.~J.~T.,  {Rappaport} S.,
  {Heger} A.,   {Pfahl} E.,  2004, \mn@doi [\apj] {10.1086/421713}, \href
  {https://ui.adsabs.harvard.edu/abs/2004ApJ...612.1044P} {612, 1044}

\bibitem[\protect\citeauthoryear{{Portegies Zwart}, {McMillan}  \&
  {Gieles}}{{Portegies Zwart} et~al.}{2010}]{2010ARA&A..48..431P}
{Portegies Zwart} S.~F.,  {McMillan} S. L.~W.,   {Gieles} M.,  2010, \mn@doi
  [\araa] {10.1146/annurev-astro-081309-130834}, \href
  {https://ui.adsabs.harvard.edu/abs/2010ARA&A..48..431P} {48, 431}

\bibitem[\protect\citeauthoryear{{Rastello}, {Iorio}, {Mapelli}, {Arca-Sedda},
  {Di Carlo}, {Escobar}, {Shenar}  \& {Torniamenti}}{{Rastello}
  et~al.}{2023}]{2023MNRAS.526..740R}
{Rastello} S.,  {Iorio} G.,  {Mapelli} M.,  {Arca-Sedda} M.,  {Di Carlo} U.~N.,
   {Escobar} G.~J.,  {Shenar} T.,   {Torniamenti} S.,  2023, \mn@doi [\mnras]
  {10.1093/mnras/stad2757}, \href
  {https://ui.adsabs.harvard.edu/abs/2023MNRAS.526..740R} {526, 740}

\bibitem[\protect\citeauthoryear{{Rivinius}, {Baade}, {Hadrava}, {Heida}  \&
  {Klement}}{{Rivinius} et~al.}{2020}]{2020A&A...637L...3R}
{Rivinius} T.,  {Baade} D.,  {Hadrava} P.,  {Heida} M.,   {Klement} R.,  2020,
  \mn@doi [\aap] {10.1051/0004-6361/202038020}, \href
  {https://ui.adsabs.harvard.edu/abs/2020A&A...637L...3R} {637, L3}

\bibitem[\protect\citeauthoryear{{Rowan}, {Thompson}, {Jayasinghe}, {Kochanek}
  \& {Stanek}}{{Rowan} et~al.}{2024}]{2024arXiv240109531R}
{Rowan} D.~M.,  {Thompson} T.~A.,  {Jayasinghe} T.,  {Kochanek} C.~S.,
  {Stanek} K.~Z.,  2024, \mn@doi [arXiv e-prints] {10.48550/arXiv.2401.09531},
  \href {https://ui.adsabs.harvard.edu/abs/2024arXiv240109531R} {p.
  arXiv:2401.09531}

\bibitem[\protect\citeauthoryear{Ruiter}{Ruiter}{2019}]{Ruiter_2019}
Ruiter A.~J.,  2019, \mn@doi [Proceedings of the International Astronomical
  Union] {10.1017/S1743921320000587}, 15, 1–15

\bibitem[\protect\citeauthoryear{{Sahu} et~al.,}{{Sahu}
  et~al.}{2022}]{2022ApJ...933...83S}
{Sahu} K.~C.,  et~al., 2022, \mn@doi [\apj] {10.3847/1538-4357/ac739e}, \href
  {https://ui.adsabs.harvard.edu/abs/2022ApJ...933...83S} {933, 83}

\bibitem[\protect\citeauthoryear{{Sana} et~al.,}{{Sana}
  et~al.}{2012}]{2012Sci...337..444S}
{Sana} H.,  et~al., 2012, \mn@doi [Science] {10.1126/science.1223344}, \href
  {https://ui.adsabs.harvard.edu/abs/2012Sci...337..444S} {337, 444}

\bibitem[\protect\citeauthoryear{{Saracino} et~al.,}{{Saracino}
  et~al.}{2022}]{2022MNRAS.511.2914S}
{Saracino} S.,  et~al., 2022, \mn@doi [\mnras] {10.1093/mnras/stab3159}, \href
  {https://ui.adsabs.harvard.edu/abs/2022MNRAS.511.2914S} {511, 2914}

\bibitem[\protect\citeauthoryear{{Sato}, {Nakasato}, {Tanikawa}, {Nomoto},
  {Maeda}  \& {Hachisu}}{{Sato} et~al.}{2015}]{2015ApJ...807..105S}
{Sato} Y.,  {Nakasato} N.,  {Tanikawa} A.,  {Nomoto} K.,  {Maeda} K.,
  {Hachisu} I.,  2015, \mn@doi [\apj] {10.1088/0004-637X/807/1/105}, \href
  {http://ads.nao.ac.jp/abs/2015ApJ...807..105S} {807, 105}

\bibitem[\protect\citeauthoryear{{Sato}, {Nakasato}, {Tanikawa}, {Nomoto},
  {Maeda}  \& {Hachisu}}{{Sato} et~al.}{2016}]{2016ApJ...821...67S}
{Sato} Y.,  {Nakasato} N.,  {Tanikawa} A.,  {Nomoto} K.,  {Maeda} K.,
  {Hachisu} I.,  2016, \mn@doi [\apj] {10.3847/0004-637X/821/1/67}, \href
  {http://ads.nao.ac.jp/abs/2016ApJ...821...67S} {821, 67}

\bibitem[\protect\citeauthoryear{{Schwab}, {Quataert}  \& {Bildsten}}{{Schwab}
  et~al.}{2015}]{2015MNRAS.453.1910S}
{Schwab} J.,  {Quataert} E.,   {Bildsten} L.,  2015, \mn@doi [\mnras]
  {10.1093/mnras/stv1804}, \href
  {https://ui.adsabs.harvard.edu/abs/2015MNRAS.453.1910S} {453, 1910}

\bibitem[\protect\citeauthoryear{{Shahaf}, {Hallakoun}, {Mazeh}, {Ben-Ami},
  {Rekhi}, {El-Badry}  \& {Toonen}}{{Shahaf}
  et~al.}{2023a}]{2023arXiv230915143S}
{Shahaf} S.,  {Hallakoun} N.,  {Mazeh} T.,  {Ben-Ami} S.,  {Rekhi} P.,
  {El-Badry} K.,   {Toonen} S.,  2023a, \mn@doi [arXiv e-prints]
  {10.48550/arXiv.2309.15143}, \href
  {https://ui.adsabs.harvard.edu/abs/2023arXiv230915143S} {p. arXiv:2309.15143}

\bibitem[\protect\citeauthoryear{{Shahaf}, {Bashi}, {Mazeh}, {Faigler},
  {Arenou}, {El-Badry}  \& {Rix}}{{Shahaf} et~al.}{2023b}]{2023MNRAS.518.2991S}
{Shahaf} S.,  {Bashi} D.,  {Mazeh} T.,  {Faigler} S.,  {Arenou} F.,  {El-Badry}
  K.,   {Rix} H.~W.,  2023b, \mn@doi [\mnras] {10.1093/mnras/stac3290}, \href
  {https://ui.adsabs.harvard.edu/abs/2023MNRAS.518.2991S} {518, 2991}

\bibitem[\protect\citeauthoryear{{Shen} et~al.,}{{Shen}
  et~al.}{2018}]{2018ApJ...865...15S}
{Shen} K.~J.,  et~al., 2018, \mn@doi [\apj] {10.3847/1538-4357/aad55b}, \href
  {http://adsabs.harvard.edu/abs/2018ApJ...865...15S} {865, 15}

\bibitem[\protect\citeauthoryear{{Shenar} et~al.,}{{Shenar}
  et~al.}{2022}]{2022NatAs...6.1085S}
{Shenar} T.,  et~al., 2022, \mn@doi [Nature Astronomy]
  {10.1038/s41550-022-01730-y}, \href
  {https://ui.adsabs.harvard.edu/abs/2022NatAs...6.1085S} {6, 1085}

\bibitem[\protect\citeauthoryear{{Shikauchi}, {Kumamoto}, {Tanikawa}  \&
  {Fujii}}{{Shikauchi} et~al.}{2020}]{2020PASJ...72...45S}
{Shikauchi} M.,  {Kumamoto} J.,  {Tanikawa} A.,   {Fujii} M.~S.,  2020, \mn@doi
  [\pasj] {10.1093/pasj/psaa030}, \href
  {https://ui.adsabs.harvard.edu/abs/2020PASJ...72...45S} {72, 45}

\bibitem[\protect\citeauthoryear{{Shikauchi}, {Tsuna}, {Tanikawa}  \&
  {Kawanaka}}{{Shikauchi} et~al.}{2023}]{2023ApJ...953...52S}
{Shikauchi} M.,  {Tsuna} D.,  {Tanikawa} A.,   {Kawanaka} N.,  2023, \mn@doi
  [\apj] {10.3847/1538-4357/acd752}, \href
  {https://ui.adsabs.harvard.edu/abs/2023ApJ...953...52S} {953, 52}

\bibitem[\protect\citeauthoryear{{Spitzer}}{{Spitzer}}{1987}]{1987degc.book.....S}
{Spitzer} L.,  1987, {Dynamical evolution of globular clusters}

\bibitem[\protect\citeauthoryear{{Tanikawa}, {Nomoto}  \&
  {Nakasato}}{{Tanikawa} et~al.}{2018}]{2018ApJ...868...90T}
{Tanikawa} A.,  {Nomoto} K.,   {Nakasato} N.,  2018, \mn@doi [\apj]
  {10.3847/1538-4357/aae9ee}, \href
  {http://adsabs.harvard.edu/abs/2018ApJ...868...90T} {868, 90}

\bibitem[\protect\citeauthoryear{{Tanikawa}, {Nomoto}, {Nakasato}  \&
  {Maeda}}{{Tanikawa} et~al.}{2019}]{2019ApJ...885..103T}
{Tanikawa} A.,  {Nomoto} K.,  {Nakasato} N.,   {Maeda} K.,  2019, \mn@doi
  [\apj] {10.3847/1538-4357/ab46b6}, \href
  {https://ui.adsabs.harvard.edu/abs/2019ApJ...885..103T} {885, 103}

\bibitem[\protect\citeauthoryear{{Tanikawa}, {Hattori}, {Kawanaka}, {Kinugawa},
  {Shikauchi}  \& {Tsuna}}{{Tanikawa} et~al.}{2023}]{2023ApJ...946...79T}
{Tanikawa} A.,  {Hattori} K.,  {Kawanaka} N.,  {Kinugawa} T.,  {Shikauchi} M.,
   {Tsuna} D.,  2023, \mn@doi [\apj] {10.3847/1538-4357/acbf36}, \href
  {https://ui.adsabs.harvard.edu/abs/2023ApJ...946...79T} {946, 79}

\bibitem[\protect\citeauthoryear{{Tanikawa}, {Cary}, {Shikauchi}, {Wang}  \&
  {Fujii}}{{Tanikawa} et~al.}{2024}]{2024MNRAS.527.4031T}
{Tanikawa} A.,  {Cary} S.,  {Shikauchi} M.,  {Wang} L.,   {Fujii} M.~S.,  2024,
  \mn@doi [\mnras] {10.1093/mnras/stad3294}, \href
  {https://ui.adsabs.harvard.edu/abs/2024MNRAS.527.4031T} {527, 4031}

\bibitem[\protect\citeauthoryear{{Thompson} et~al.,}{{Thompson}
  et~al.}{2019}]{2019Sci...366..637T}
{Thompson} T.~A.,  et~al., 2019, \mn@doi [Science] {10.1126/science.aau4005},
  \href {https://ui.adsabs.harvard.edu/abs/2019Sci...366..637T} {366, 637}

\bibitem[\protect\citeauthoryear{{Toonen}, {Nelemans}  \& {Portegies
  Zwart}}{{Toonen} et~al.}{2012}]{2012A&A...546A..70T}
{Toonen} S.,  {Nelemans} G.,   {Portegies Zwart} S.,  2012, \mn@doi [\aap]
  {10.1051/0004-6361/201218966}, \href
  {https://ui.adsabs.harvard.edu/abs/2012A&A...546A..70T} {546, A70}

\bibitem[\protect\citeauthoryear{{Wang}, {Nitadori}  \& {Makino}}{{Wang}
  et~al.}{2020a}]{2020MNRAS.493.3398W}
{Wang} L.,  {Nitadori} K.,   {Makino} J.,  2020a, \mn@doi [\mnras]
  {10.1093/mnras/staa480}, \href
  {https://ui.adsabs.harvard.edu/abs/2020MNRAS.493.3398W} {493, 3398}

\bibitem[\protect\citeauthoryear{{Wang}, {Iwasawa}, {Nitadori}  \&
  {Makino}}{{Wang} et~al.}{2020b}]{2020MNRAS.497..536W}
{Wang} L.,  {Iwasawa} M.,  {Nitadori} K.,   {Makino} J.,  2020b, \mn@doi
  [\mnras] {10.1093/mnras/staa1915}, \href
  {https://ui.adsabs.harvard.edu/abs/2020MNRAS.497..536W} {497, 536}

\bibitem[\protect\citeauthoryear{{Webbink}}{{Webbink}}{1984}]{1984ApJ...277..355W}
{Webbink} R.~F.,  1984, \mn@doi [\apj] {10.1086/161701}, \href
  {http://ads.nao.ac.jp/abs/1984ApJ...277..355W} {277, 355}

\bibitem[\protect\citeauthoryear{{Woosley} \& {Baron}}{{Woosley} \&
  {Baron}}{1992}]{1992ApJ...391..228W}
{Woosley} S.~E.,  {Baron} E.,  1992, \mn@doi [\apj] {10.1086/171338}, \href
  {https://ui.adsabs.harvard.edu/abs/1992ApJ...391..228W} {391, 228}

\bibitem[\protect\citeauthoryear{{Woosley}, {Taam}  \& {Weaver}}{{Woosley}
  et~al.}{1986}]{1986ApJ...301..601W}
{Woosley} S.~E.,  {Taam} R.~E.,   {Weaver} T.~A.,  1986, \mn@doi [\apj]
  {10.1086/163926}, \href {http://adsabs.harvard.edu/abs/1986ApJ...301..601W}
  {301, 601}

\bibitem[\protect\citeauthoryear{{Yamaguchi} et~al.,}{{Yamaguchi}
  et~al.}{2024}]{2024MNRAS.527.11719}
{Yamaguchi} N.,  et~al., 2024, \mn@doi [\mnras] {10.1093/mnras/stad4005}, \href
  {https://ui.adsabs.harvard.edu/abs/2024MNRAS.527.11719} {527, 11719}

\bibitem[\protect\citeauthoryear{{Zenati}, {Perets}, {Dessart},
  {Jacobson-Gal{\'a}n}, {Toonen}  \& {Rest}}{{Zenati}
  et~al.}{2023}]{2023ApJ...944...22Z}
{Zenati} Y.,  {Perets} H.~B.,  {Dessart} L.,  {Jacobson-Gal{\'a}n} W.~V.,
  {Toonen} S.,   {Rest} A.,  2023, \mn@doi [\apj] {10.3847/1538-4357/acaf65},
  \href {https://ui.adsabs.harvard.edu/abs/2023ApJ...944...22Z} {944, 22}

\bibitem[\protect\citeauthoryear{{Zheng} et~al.,}{{Zheng}
  et~al.}{2023}]{2023SCPMA..6629512Z}
{Zheng} L.-L.,  et~al., 2023, \mn@doi [Science China Physics, Mechanics, and
  Astronomy] {10.1007/s11433-023-2247-x}, \href
  {https://ui.adsabs.harvard.edu/abs/2023SCPMA..6629512Z} {66, 129512}

\bibitem[\protect\citeauthoryear{{van den Heuvel}}{{van den
  Heuvel}}{2007}]{2007AIPC..924..598V}
{van den Heuvel} E.~P.~J.,  2007, in {di Salvo} T.,  {Israel} G.~L.,
  {Piersant} L.,  {Burderi} L.,  {Matt} G.,  {Tornambe} A.,   {Menna} M.~T.,
  eds,  American Institute of Physics Conference Series Vol. 924, The
  Multicolored Landscape of Compact Objects and Their Explosive Origins. pp
  598--606 (\mn@eprint {arXiv} {0704.1215}), \mn@doi{10.1063/1.2774916}

\bibitem[\protect\citeauthoryear{{van den Heuvel} \& {Tauris}}{{van den Heuvel}
  \& {Tauris}}{2020}]{2020Sci...368.3282V}
{van den Heuvel} E. P.~J.,  {Tauris} T.~M.,  2020, \mn@doi [Science]
  {10.1126/science.aba3282}, \href
  {https://ui.adsabs.harvard.edu/abs/2020Sci...368.3282V} {368, eaba3282}

\makeatother
\end{thebibliography}
\end{document}